\documentclass[a4paper,12pt]{article}
\usepackage{epsfig,amssymb,amsmath}
\usepackage{jheppub} 
\usepackage{psfrag}
\usepackage{esint} 
\usepackage{breqn}

\def \tr {\mathop{\rm tr}\nolimits}

\def \e  {\mathop{\rm e}\nolimits}

\newcommand\lr[1]{{\left({#1}\right)}}
\newcommand \widebar [1] {\overline{#1}}

\newcommand \vev [1] {\langle{#1}\rangle}

\newcommand \ket [1] {|{#1}\rangle}
\newcommand \bra [1] {\langle {#1}|}
\newcommand\re[1]{(\ref{#1})}
\def \qqquad {\qquad\quad}
\def \qqqquad {\qquad\qquad}

\def\numberbysection{\@addtoreset{equation}{section}
                     \def\theequation{\thesection.\arabic{equation}}}

\preprint{\small  \parbox[t]{25mm}{IPhT-T17/053}}

\title{\Large Instanton effects in correlation functions on the light-cone}

\author{G.P.~Korchemsky}

\affiliation{Institut de Physique Th\'eorique\footnote{Unit\'e Mixte de Recherche 3681 du CNRS}, Universit\'e Paris Saclay, CNRS, CEA, 91191 Gif-sur-Yvette}

\abstract{We study instanton corrections to four-point correlation correlation function of half-BPS operators in $\mathcal N=4$ SYM 
in the light-cone limit when operators become null separated in a sequential manner. We exploit the relation
between the correlation function in this limit  and light-like rectangular Wilson loop to determine the leading instanton contribution to the former
from the semiclassical result for the latter.
We verify that the light-like rectangular Wilson loop  satisfies anomalous conformal Ward identities 
nonperturbatively, in the presence of instantons. We then use these identities to compute the leading instanton contribution to 
the light-like cusp anomalous dimension and to 
anomalous dimension of twist-two operators
with large spin.
} 

\begin{document}

\maketitle

\flushbottom
 
\section{Introduction}

Maximally supersymmetric Yang-Mills theory ($\mathcal N=4$ SYM) possesses a remarkable electric-magnetic duality, also 
known as $S-$duality \cite{Montonen:1977sn,Witten:1978mh,Osborn:1979tq}. It establishes the equivalence between certain correlation functions computed at weak and at strong coupling. 
Testing the $S-$duality proves to be a complicated task as it requires understanding these functions
at strong coupling \cite{Gomis:2009ir,Gomis:2009xg}.  

In the case of  $\mathcal N=4$ SYM with the $SU(N)$ gauge group, the $S-$duality predicts that the correlation functions
of half-BPS operators $G_n=\vev{O_{\mathbf 20'}(x_1)\dots O_{\mathbf 20'}(x_n)}$ should be invariant under modular $SL(2,\mathbb Z)$ transformations acting on the complexified coupling
constant $\tau = \theta/(2\pi) + 4\pi i/g^2$. Two- and three-point correlation functions of half-BPS operators are protected from
quantum corrections and trivially verify the $S-$duality. For higher number of points, the functions $G_n$ receive quantum corrections and have a nontrivial dependence on the coupling constant. In order to test the $S-$duality, we have to find perturbative contribution
to $G_n$ for finite $N$ and supplement it with instanton corrections. 
Although these corrections are exponentially small at large $N$, they play a crucial role in restoring the $S-$duality.
 
In this paper we study instanton effects in four-point correlation function of half-BPS operators. Quantum corrections to 
$G_4$ are described by a single function $\mathcal G(u,v)$ of two
cross ratios $u=x_{12}^2x_{34}^2/(x_{13}^2x_{24}^2)$ and  $v=x_{23}^2x_{14}^2/(x_{13}^2x_{24}^2)$ (with $x_{ij}^2=(x_i-x_j)^2$). It has the following general form at weak coupling
\begin{align}\label{G}
\mathcal G(u,v) =  \Phi_0(u,v;g^2) + \sum_{n \ge 1} \Big(\e^{2\pi i n \tau}  + \e^{-2\pi i n \bar \tau} \Big) \Phi_{n}(u,v;g^2)\,,
\end{align}
where the first term $\Phi_0(u,v;g^2)$ is a perturbative correction and 
the second one is a nonperturbative correction due to  $n$ (anti) instantons. The function $\Phi_{n}(u,v)$ describes 
the contribution of quantum fluctuations of instantons and runs in powers of $g^2$. 
\footnote{In general, the functions $\Phi_0$ and $\Phi_n$ could also receive $O(\e^{-16\pi^2/g^2})$  corrections due to 
instanton-antiinstanton configurations, but their status remains unclear. }

At present, the instanton corrections to \re{G} are known to the lowest order in $g^2$. 
The corresponding function $\Phi_n^{(0)} =\Phi_{n}(u,v;0)$
can be found in the semiclassical approximation following the standard approach (for a review, see \cite{Belitsky:2000ws,Dorey:2002ik,Bianchi:2007ft}). In this approximation 
the quantum fluctuations are frozen and the correlation function is given by a 
finite-dimensional integral over the collective coordinates of instantons. 
An explicit 
expression for the function $\Phi_{n}^{(0)}(u,v)$ is known in one-instanton sector ($n=1$) as well as for an arbitrary number of instantons $n$ in the large $N$ limit.
To go beyond the semiclassical approximation, we have to include quantum fluctuations of instantons. Their contribution 
to $\Phi_n(u,v;g^2)$ 
scales as $O(g^{2\ell})$ where integer positive $\ell$ counts the number of instanton loops.
It is much more difficult to compute such corrections and a little progress has been made over the last decade.

Perturbative corrections to \re{G} are known to have some additional structure \cite{Eden:2011we,Eden:2012tu} which allows us to 
construct integral representation for the function $\Phi_0(u,v;g^2)$ to any order in $g^2$ without going through a
Feynman diagram calculation. Moreover,  $\mathcal N=4$ SYM is believed to be integrable in the planar limit \cite{Beisert:2010jr}. In 
application to the correlation function \re{G}, this opens up the possibility to determine perturbative contribution $\Phi_0(u,v;g^2)$ for arbitrary  
't Hooft coupling at large $N$ \cite{Eden:2016xvg,Fleury:2016ykk}. A natural question is whether some of these remarkable properties survive
in \re{G} in the instanton sector. 

As a first example, we examine behaviour of the correlation function $G_4$ at 
short distances $x_{12}^2\to 0$, or equivalently for $u\to 0$ and $v\to 1$. As follows from the OPE, the leading contribution 
to \re{G} in this limit, $\mathcal G(u,v)\sim C_{K}^2\, u^{\gamma_K/2}$, comes from Konishi operator, unprotected operator with smallest scalling dimension $\Delta=2+\gamma_K$. The anomalous dimension of this operator, 
$\gamma_K$, and its structure constant in the OPE of two half-BPS operators, $C_K$,  have expansion at weak coupling
similar to \re{G}. The leading instanton correction to $\gamma_K$ and $C_K$ can be found by examining the asymptotic
behaviour of the second term on the right-hand side of  \re{G}. In this way one obtains, using the known results for the function
$G_4$ in the semiclassical approximation, that $\gamma_K$ and $C_K$ do not receive  
$O(\e^{2\pi i  \tau})$ instanton correction \cite{Bianchi:1999ge,Arutyunov:2000im,Bianchi:2001cm,Kovacs:2003rt}. 
Thus, the leading corrections to $\gamma_K$ and $C_K$ can only come from quantum instanton corrections to \re{G}. 

The same quantities 
can be also extracted from the two- and three-point correlation functions, $\vev{K(x) K(0)}\sim 1/(x^2)^{2+\gamma_K}$
and $C_K\sim \vev{O_{\mathbf 20'}O_{\mathbf 20'} K}$, respectively. Computation of these correlation
functions in the semiclassical approximation yields the following result for the leading instanton corrections $C_K^{(\rm inst)} = O(g^2 \e^{2\pi i  \tau})$ and
$\gamma_K^{(\rm inst)} = O(g^4 \e^{2\pi i \tau})$ (explicit expressions can be found in \cite{Alday:2016tll}). 
Notice that both expressions have additional factors of $g^2$ as compared with the semiclassical  $O(\e^{2\pi i \tau})$
contribution to \re{G}.
To get the same expressions for  $C_K$ and $\gamma_K$ from the four-point correlation function $G_4$, one would
have to take into account one- and two-loop instanton corrections to \re{G}, respectively. 

This example illustrates a hidden simplicity of instanton effects -- finding the leading quantum instanton contribution
to the four-point correlation function at short distances, $\mathcal G(u,v)\sim C_{K}^2\, u^{\gamma_K/2}$, can be mapped into a semiclassical calculation of two- and three-point 
correlation functions of the Konishi operator \cite{Alday:2016tll}. 

We show in this paper that analogous phenomenon also happens for $G_4$ in the light-like limit
$x_{i,i+1}^2\to 0$ (with $x_{i+4} \equiv x_i$) when four half-BPS operators become light-like separated in a sequential manner.
In this limit, the correlation function is expected to have the following form \cite{Alday:2010zy}
\begin{align}\label{G-lim}
G_4 = {1\over x_{12}^2 x_{23}^2 x_{34}^2 x_{41}^2} \mathcal G(u,v) \,,
\end{align}
where the product of four scalar propagators defines the leading asymptotic behavior and the function $\mathcal G(u,v)$
is given by \re{G} for $u,v\to 0$. At weak coupling, perturbative corrections to \re{G} are enhanced by powers of logarithms of $u$ and $v$.
Such corrections can be summed to all orders in $g^2$ leading to \cite{Alday:2010zy,Alday:2013cwa}
\begin{align}\label{G-lc}
\mathcal G(u,v) = W_4 \times J = \exp\bigg[{-\frac12 \Gamma_{\rm cusp}(g^2) \ln u \ln v + \dots}\bigg]\,.
\end{align} 
Here dots denote subleading corrections and $\Gamma_{\rm cusp}(g^2)$ is the light-like cusp anomalous dimension in the
adjoint representation of the $SU(N)$. 

The same anomalous dimension controls divergences of cusped light-like Wilson
loops and its appearance in \re{G-lc} is not accidental. As was shown in \cite{Alday:2010zy}, the leading asymptotics of the function 
$\mathcal G(u,v)$ is described by (an appropriately regularized) rectangular light-like Wilson loop  
\begin{align}\label{W4-A}
W_4 = \vev{0|\,\tr_A P \exp\lr{ig \oint_{C_4} dx\cdot A(x)}|0}\,,
\end{align}
evaluated along light-like rectangle $C_4$ with vertices at points $x_i$. The subscript $\scriptstyle A $ indicates that $W_4$ is defined in the adjoint representation of the $SU(N)$.
The subleading (logarithmically enhanced) corrections to \re{G-lc} come from the
so-called jet factor $J$. Its form is fixed by the crossing symmetry of the four-point correlation function \cite{Alday:2013cwa}. 

In this paper, we compute the leading instanton correction to the four-point correlation function \re{G} in the light-cone limit $x_{i,i+1}^2\to 0$. Notice that this limit
is Minkowskian in nature whereas instantons are defined in Euclidean signature. To find 
instanton corrections to \re{G}, we shall determine the function $\mathcal G(u,v)$ in Euclidean domain of 
$u$ and $v$ and, then, analytically continue it to $u, v\to 0$. 

As in the previous example, we start with the semiclassical approximation to \re{G}. As was  shown in  \cite{Bianchi:2013xsa},
the instanton corrections to $\mathcal G(u,v)$   scale in this approximation as $O(uv)$ and, therefore, they do not modify the asymptotic 
behaviour \re{G-lc}. 
To go beyond the semiclassical approximation, we analyze the light-cone asymptotics of $G_4$ and argue that the relation
between $\mathcal G(u,v)$ and light-like rectangular Wilson loop mentioned above also holds in the presence of instantons. 
This relation allows us to establish the correspondence between the leading (quantum) instanton correction to $\mathcal G(u,v)$ 
and the semiclassical result for $W_4$. 

We show that the resulting expression for $\mathcal G(u,v)$
takes the same form as in perturbation theory \re{G-lc} with the important difference that the light-like
cusp anomalous dimension in \re{G-lc} is modified by the instanton correction. In the simplest case of the $SU(2)$ gauge group, 
this correction  in one-(anti)instanton sector  is given by
\begin{align}\label{cusp-inst}
 \Gamma_{\rm cusp}(g^2) = - \frac4{15}  \lr{g^2\over 4\pi^2}^4   \lr{ \e^{2\pi i \tau} +\e^{-2\pi i \bar\tau} }\,.
\end{align}
Following \cite{Dorey:1998xe,Dorey:1999pd,Dorey:2002ik}, this result can be generalized to the $SU(N)$  gauge group and to the case of multi-instantons 
at large $N$. To obtain the same result \re{cusp-inst} from the direct calculation of the four-point correlation function, 
one would have to compute quantum instanton corrections to \re{G} at order $O(g^8)$.

As a byproduct of our analysis, we verify that the light-like rectangular Wilson loop $W_4$ satisfies the anomalous conformal Ward identities 
\cite{Drummond:2007au}
nonperturbatively, in the presence of instantons. We also determine the leading instanton correction to anomalous dimension
$\gamma_S$ of twist-two operators
with large spin $S\gg 1$. This anomalous dimension scales logarithmically with the spin (see Eq.~\re{gamma-S} below) and its behaviour is controllled by the cusp anomalous dimension \cite{Korchemsky:1988si}. Making use of \re{cusp-inst} we find that the leading  instanton contribution 
scales as $\gamma_S \sim g^8 \e^{2\pi i\tau}\ln S$. This agrees with the finding of \cite{Alday:2016jeo} that $\gamma_S$ does not receive 
$O(g^4 \e^{2\pi i\tau})$ correction for any spin $S>2$.

The paper is organized as follows. In Section 2 we analyze asymptotic behaviour of the four-point correlation function in the
light-cone limit $x_{i,i+1}^2\to 0$ in the presence of instantons and discuss its relation with the light-like Wilson loop.
In Section 3 we compute instanton contribution to the light-like rectangular Wilson loop in the semiclassical approximation. We then use it  in Section~4 to determine the leading instanton correction to the cusp anomalous dimension. Section~5 contains
concluding remarks. Details of the calculation are presented in four appendices.

\section{Correlation functions in the light-cone limit}

In this section we examine instanton corrections to a four-point correlation function of scalar half-BPS operators  
\begin{align}\label{O20}
O_{\bf 20'}(x) = Y_{AB} Y_{CD} \tr[ \phi^{AB} \phi^{CD}(x)]\,.
\end{align}
Here auxiliary $SU(4)$ tensors satisfy $\epsilon^{ABCD}Y_{AB} Y_{CD} =0$ and serve to project the operator onto
representation $\bf 20'$ of the $SU(4)$. In virtue of $\mathcal N=4$ superconformal symmetry, the dependence 
of the four-point correlation function 
\begin{align}\label{G4-def}
G_4=\vev{O_{\bf 20'}(x_1)O_{\bf 20'}(x_2)O_{\bf 20'}(x_3)O_{\bf 20'}(x_4)} 
\end{align}
on $Y-$variables can be factored into a universal kinematical factor independent on the coupling constant \cite{Eden:2000bk}. In what follows we do not display
this factor and concentrate on the dynamical part $\mathcal G(u,v)$ that depends on the cross ratios only. 

In the Born approximation, $G_4$ reduces to the sum of terms each given by the product of free scalar propagators. In
the light-like limit, $x_{i,i+1}^2\to 0$, the leading contribution to $G_4$ comes from only one term of the form \re{G-lim}
with $\mathcal G(u,v) =1$.~\footnote{In what follows we discard a disconnected part of the correlation function.} Going beyond this
approximation, we apply the OPE to each pair of neighbouring operators in \re{G4-def}, e.g. 
\begin{align}\label{OPE}
O_{\bf 20'}(x_1)O_{\bf 20'}(x_2) =  \sum_S {C_S \over (x_{12}^2)^{2-t_S/2}}x_{12}^{\mu_1}\dots x_{12}^{\mu_S} O_{\mu_1\dots \mu_S} (x_1)\,,
\end{align}
where the sum runs over local operators with Lorentz spin $S$, dimension $\Delta_S$ and twist $t_S=\Delta_S -S\ge 2$. 
For $x_{12}^2\to 0$ the dominant contribution to \re{OPE}  comes from twist-two operators with arbitrary spin $S$ and
scaling dimension $\Delta_S= 2 + S + \gamma_S$.~\footnote{Strictly speaking, this is true only if the anomalous dimension of the operator is small, or equivalently the value of its twist is close to that
in a free theory. If the anomalous dimension of twist-two operator is large  $\gamma_S\sim 2$, it collides with the 
twist-four operators.} 
It scales as $O(1/x_{12}^2)$ and yields the expected asymptotic behaviour \re{G-lim}. In the similar manner, the remaining $O(1/x_{i,i+1}^2)$
factors in \re{G-lim} come from the twist-two operators propagating in other OPE channels. 

A detailed analysis shows \cite{Alday:2010zy,Alday:2013cwa}, that the
leading asymptotic behaviour of the function $\mathcal G(u,v)$ for $u,v\to 0$ is governed by twist-two operators with large spin $S=O(1/\sqrt{u})$ or $S=O(1/\sqrt{v})$ depending on the OPE channel. The anomalous dimension of such operators grows logarithmically with the
spin, 
\begin{align}\label{gamma-S}
\gamma_S = 2 \Gamma_{\rm cusp} \ln S +O(S^0)\,,
\end{align}
and generates corrections to $\mathcal G(u,v)$ enhanced by powers of $\ln u$
and $\ln v$. As we see in a moment, this observation simplifies the calculation of instanton corrections to \re{G-lim}.
  
Instantons are classical configurations of fields (scalar, gaugino and gauge fields) satisfying 
equations of motion in Euclidean $\mathcal N=4$ SYM \cite{Belavin:1975fg}. To compute their contribution to the correlation function \re{G4-def} at weak coupling, we have to 
go through few steps. First, we decompose all fields (that we denote generically by $\Phi(x)$) into classical, instanton part and quantum fluctuations, 
\begin{align}\label{dec}
\Phi(x) = {1\over g} \Phi_{\rm inst}(x) + \Phi_q(x)\,.
\end{align}
Here we introduced the factor of $1/g$ to emphasize that $\Phi_{\rm inst}(x)$ does not depend on the coupling constant. Then, we substitute 
\re{dec} into \re{G4-def} and integrate over quantum fluctuations $\Phi_q$ and over collective coordinates of instantons.
Finally, we match the resulting expression for $G_4$ into \re{G-lim}, identify the function $\mathcal G(u,v)$ and analytically
continue it to small $u$ and $v$.

\begin{figure}
\psfrag{x1}[cc][cc]{$\scriptstyle x_1$}\psfrag{x2}[cc][cc]{$\scriptstyle x_2$}
\psfrag{x3}[cc][cc]{$\scriptstyle x_3$}\psfrag{x4}[cc][cc]{$\scriptstyle x_4$}
\psfrag{I}[cc][cc]{$I$}
\psfrag{(a)}[cc][cc]{(a)}\psfrag{(b)}[cc][cc]{(b)}\psfrag{(c)}[cc][cc]{(c)}\psfrag{(d)}[cc][cc]{(d)}
\includegraphics[width = \textwidth]{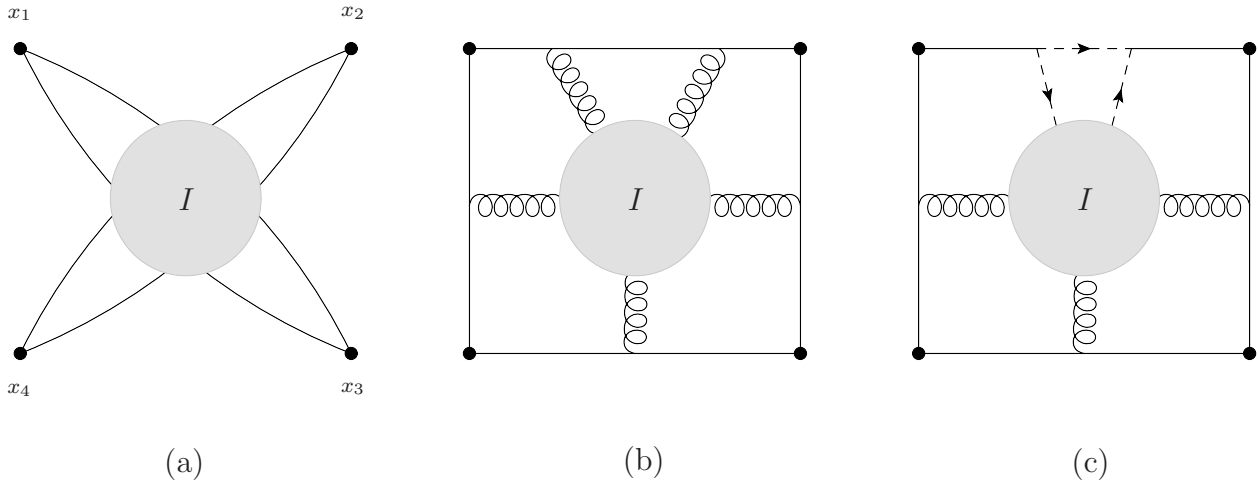}
\caption{Instanton corrections to the four-point correlation function: (a) contribution in the semiclassical approximation; (b) leading contribution in the
light-cone limit; (c) example of subleading contribution. Solid, wavy and dashes lines denote scalars, gauge fields and gauginos, respectively.  
Lines attached to the central blob represent the instanton background.
}\label{fig1}
\end{figure}

In the semiclassical approximation, we neglect quantum fluctuations in \re{dec} and obtain the following expression for $G_4$ 
\begin{align}\label{G4-semi}
G_{4, \rm inst}= \int d\mu_{\rm phys} \e^{-S_{\rm phys}} O_{\bf 20'}(x_1)O_{\bf 20'}(x_2)O_{\bf 20'}(x_3)O_{\bf 20'}(x_4)\,,
\end{align}
where the half-BPS operators \re{O20} are replaced by their expressions in the instanton background
and are integrated over the collective coordinates of the instantons. The relation  \re{G4-semi} can be represented diagrammatically as shown in Figure~\ref{fig1}(a). 

For the one-instanton configuration in $\mathcal N=4$
SYM with the $SU(2)$ gauge group the integration measure is given by \cite{Bianchi:1998nk}
\begin{align}\label{measure}
\int d\mu_{\rm phys} \e^{-S_{\rm phys}} = {g^8 \over 2^{34} \pi^{10}} \e^{2\pi i\tau} \int d^4 x_0 \int{d\rho\over \rho^5} \int d^8 \xi
\int d^8 \bar\eta\,.
\end{align}
Here bosonic collective coordinates $x_0^\mu$ and $\rho$ define the position of the instanton and its size, respectively. 
Fermionic coordinates $\xi_\alpha^A$ and $\bar\eta_{\dot\alpha}^A$ (with $A=1,\dots,4$ and $\alpha,\dot\alpha=1,2$)
reflect the invariance of $\mathcal N=4$ SYM under superconformal  transformations. 
For the correlation function \re{G4-semi} to be different from zero, the product of four
half-BPS operators in \re{G4-semi} should soak all $16$ fermion modes. 

For the $SU(N)$
gauge group the instanton $\Phi_{\rm inst}(x)$ depends on the additional bosonic and fermion modes. In what follows we shall concentrate on the $SU(2)$ case and discuss generalization to the $SU(N)$
later in Section~\ref{sect:res}.

In $\mathcal N=4$ SYM with the $SU(2)$ gauge group, the one-instanton solution can be obtained  \cite{Zumino:1977yh} by applying  
superconformal transformations $\exp(\xi Q+\bar\eta \bar S)$ to the special field configuration consisting of vanishing 
scalar and gaugino fields  and gauge field given by the celebrated BPST instanton.~\footnote{This field configuration is annihilated by the remaining $\bar Q$ and $S$ generators.} For gauge field this
leads to 
\begin{align}\label{A-exp}
{}& A_{\rm inst}(x) = A^{(0)} +A^{(4)} + A^{(8)}  \,, 
\end{align}
where $A^{(0)}$ is the BPST instanton \cite{Belavin:1975fg} and $A^{(4n)}$ denotes component containing $4n$ fermion modes $\xi_\alpha^A$ and $\bar\eta_{\dot\alpha}^A$. In virtue of the $SU(4)$ symmetry, each subsequent term of the expansion has four modes more.
Expressions for the scalar and gaugino fields have a form similar to \re{A-exp} with the only difference that the lowest term of the expansion has a nonzero number of fermion modes whose value is dictated by the $R-$charge of the fields. Notice that the expansion \re{A-exp} is shorter than one might expect as the $SU(4)$ symmetry allows for the presence of terms with up to $16$ fermion modes.
It turns out, however, that all field components with the number of fermion modes
exceeding $8$ vanish due to $\mathcal N=4$ superconformal symmetry \cite{Alday:2016jeo}. The explicit expressions for various components in \re{A-exp}
can be found in \cite{Belitsky:2000ws,Alday:2016jeo}.

We recall that relations \re{A-exp} define the classical part of \re{dec}. Replacing the scalar field in the definition of the half-BPS operator 
\re{O20} with $\phi(x)=g^{-1} \phi_{\rm inst}(x)$ we find that $O_{\bf 20'}(x)$ scales in the instanton background as 
$O(1/g^2)$ and contains $4$ fermion modes. Then, we can apply \re{G4-semi} and \re{measure} to arrive at the following result for
the four-point correlation function in the semiclassical approximation \cite{Bianchi:1998nk}
\begin{align}\label{G4-bare}
G_{4, \rm inst} =  {108\over\pi^2}  \e^{2\pi i\tau}  {u v \bar D_{44}(u,v)\over x_{12}^2 x_{23}^2 x_{34}^2 x_{14}^2}\,,
\end{align}
where $\bar D-$function is defined in Appendix~\ref{app:D-fun}. The contribution of anti-instanton is given by a complex conjugated
expression.
The relation \re{G4-bare} holds for arbitrary $x_{i,i+1}^2$.
For $x_{i,i+1}^2\to 0$ we find from \re{G-lim}  that the instanton contribution vanishes as
$\mathcal G(u,v) = O(u v)$ for $u,v\to 0$. Thus, the one-instanton correction \re{G4-bare} does not modify  
asymptotic behaviour of $G_4$ in the light-cone limit \cite{Bianchi:2013xsa}. 

The same result can be obtained using the OPE. In the semiclassical approximation, the product of the 
operators on the left-hand side of \re{OPE} reduces to the product of two functions describing the classical profile of half-BPS operators. It is
obviously regular for $x_{12}^2\to 0$ and, therefore, cannot produce $1/x_{12}^2$ singularity that is needed to get a finite result
for $\mathcal G(u,v)$ in \re{G-lim}. For such singularity to arise, we have to go beyond the semiclassical approximation in \re{dec} and
exchange quantum fluctuations between the two operators in \re{OPE}. To lowest order in the coupling we have
\begin{align}
\vev{\Phi(x_1)\Phi(x_2)} = {1\over g^2} \Phi_{\rm inst}(x_1)\Phi_{\rm inst}(x_2) + \vev{\Phi_q(x_1)\Phi_q(x_2)}\,,
\end{align} 
where $\vev{\Phi_q(x_1)\Phi_q(x_2)}\sim 1/x_{12}^2$. Notice that the quantum fluctuation produces $1/x_{12}^2$ singularity but
its contribution is suppressed by the factor of $g^2$ as compared with the semiclassical result.
 
For the correlation function to have the expected form \re{G-lim} with nonvanishing $\mathcal G(u,v)$, at least one 
quantum fluctuation has to be exchanged between each pair of neighboring operators in \re{G4-def}. As follows from the above
analysis, the corresponding contribution to $\mathcal G(u,v)$ has the following dependence on the coupling constant
\begin{align}\label{G8}
\mathcal G(u,v) =  O(g^8 \e^{2\pi i\tau})\,,
\end{align}
where each fluctuation brings in the factor of $g^2$. Comparing this relation with \re{G} we find that the first three
terms in the  expansion of the instanton induced function $\Phi_n(u,v;g^2)$ in powers of $g^2$ should vanish in the light-cone limit $u,v\to 0$, in the one-instanton sector at least.

To the leading order in $g^2$, the dominant contribution to \re{G8} comes 
from Feynman diagrams shown in Figure~\ref{fig1}(b). They contain four scalar propagators connecting the points
$x_i$ and $x_{i+1}$. In the first-quantized picture, these diagrams describe a scalar particle propagating between
the points $x_1,\dots,x_4$ in an external 
instanton gauge field. Notice that the particle can also interact with 
instanton fields of gaugino and scalar but this leads to a subleading contribution. To show this, consider the diagram shown 
in Figure~\ref{fig1}(c). It contains two Yukawa vertices 
and its contribution to $G_4$ has the same dependence on the coupling constant as \re{G8}. However, in distinction from the diagram
shown in Figure~\ref{fig1}(b), it does not produce $1/x_{12}^2$ singularity. Indeed, as follows from \re{OPE}, the leading
behaviour  $G_4\sim 1/(x_{12}^2)^{2-t/2}$ is controlled by the twist of exchanged operators. For the diagram shown in Figure~\ref{fig1}(c)
such operators are built from two scalar and two gaugino fields and their twist satisfies $t\ge 4$. For
the diagram shown in Figure~\ref{fig1}(b), the leading operators have twist two  and are of
a schematic form 
$\tr[\Phi D_+^S \Phi]$, where $D_+=\partial_+ + ig A_+$ is a light-cone component of the covariant derivative. 

Thus, the leading contribution to $G_4$ for $x_{i,i+1}^2\to 0$ only comes from diagrams
shown in Figure~\ref{fig1}(b).  Denoting the scalar propagator in the instanton background  as $D(x_i,x_{i+1})$, we obtain the following result for the correlation
function in the light-like limit  
\begin{align}\label{G4=D4}
G_{4, \rm inst}= \vev{\tr\left[ D(x_1,x_2)D(x_2,x_3)D(x_3,x_4)D(x_4,x_1)\right]}_{\rm inst}\,,
\end{align}
where $\vev{\dots}_{\rm inst}$ denotes integration over the collective coordinates of instantons with the measure
\re{measure}. This result is rather general and it holds for multi-instanton contribution to $G_4$ in $\mathcal N=4$ SYM with an arbitrary gauge group.

We can argue following \cite{Alday:2010zy} that the relation \re{G4=D4} leads to the same factorized expression
\re{G-lc} for the function $\mathcal G(u,v)$ as in perturbation theory. 
The propagator $D(x_1,x_2)$
depends on two momentum scales,  $1/x_{12}^2$ and $1/\rho^2$, which define a proper energy of the scalar particle and its
interaction energy with the instanton background, respectively. For $1/x_{12}^2\gg 1/\rho^2$, or equivalently
$x_{12}^2\ll\rho^2$, the instanton carries small energy and its interaction with the scalar particle can be treated semiclassically. 
In this limit, $D(x_1,x_2)$ reduces to a free scalar propagator multiplied by the eikonal phase given by the
Wilson line evaluated along the light-cone segment $[x_1,x_2]$. Taking the product of four Wilson lines corresponding to
four propagators in \re{G4=D4}, we obtain that the contribution of instanton with $x_{i,i+1}^2 \ll \rho^2$ is described by 
the rectangular light-like Wilson loop defined in \re{W4-A}. The gauge fields in \re{W4-A} are now replaced by the instanton solution \re{A-exp} and integration over its moduli
is performed with the measure \re{measure}. For $x_{i,i+1}^2\gg\rho^2$ the eikonal approximation is not applicable. The contribution from this region, denoted by $J$ in \re{G-lc}, can be determined from the crossing symmetry of $G_4$
in the same way as it was done in perturbation theory \cite{Alday:2013cwa}.

\section{Light-like Wilson loop in the instanton background}

We demonstrated in the previous section that the leading light-cone asymptotics of the correlation function $G_4$ is described by 
light-like rectangular Wilson loop $W_4$. 
Due to the presence of cusps on the integration contour, $W_4$ develops specific ultraviolet divergences. In the expression
for the correlation function \re{G-lc}, these divergences cancel against those of the $J-$function in such a way that the UV cut-off
of the Wilson loop is effectively replaced with $\mu^2\sim 1/x_{i,i+1}^2$.

\subsection{Conformal Ward identities}\label{sect:conf}

Let us start with summarizing the properties of $W_4$. As was shown in \cite{Drummond:2007au}, the conformal symmetry 
restricts the dependence of $W_4$ on kinematical invariants 
\begin{align}\label{ZZF}
W_4 = Z(x_{13}^2\mu^2)Z(x_{24}^2\mu^2) F_4(x_{13}^2/x_{24}^2)\,,
\end{align}
where $Z$ and $F_4$ are the divergent and finite parts, respectively. The dependence of $W_4$ on the UV cut-off $\mu^2$
is described by the evolution equation
\begin{align}\label{Z}
\lr{\mu^2 {\partial\over \partial\mu^2}}^2 \ln W_4 = - 2\Gamma_{\rm cusp}(g^2) \,,
\end{align}
where the light-like cusp anomalous dimension $\Gamma_{\rm cusp}(g^2)$ depends on the representation of the $SU(N)$ 
gauge group in which the Wilson loop is defined. The general solution to this equation depends on the so-called collinear anomalous
dimension $\Gamma_{\rm col}(g^2)$. It appears as a coefficient in front of $\ln(x_{i,i+2}^2\mu^2)$ in the expression for $\ln W_4$ and depends on 
the choice of the regularization.
The finite part of \re{ZZF} is uniquely fixed by the conformal symmetry  
\footnote{Since the Wilson loop \re{ZZF} is defined in the adjoint
representation, this expression is the square of the one found in \cite{Drummond:2007au}.}
\begin{align}\label{F4}
F_4= \exp\left[\frac12 \Gamma_{\rm cusp}(g^2)\ln^2 (x_{13}^2/x_{24}^2) \right]\,.
\end{align} 
Combining together \re{ZZF} and \re{F4} we obtain the following relation for $W_4$
\begin{align}\label{dd}
{\partial \over \partial  \ln x_{13}^2}{\partial \over \partial \ln x_{24}^2} \ln W_4 = - \Gamma_{\rm cusp}(g^2) \,,
\end{align}
where the dependence on the UV cut-off disappears 
since the second derivative annihilates the divergent part of $\ln W_4$. This relation allows us to 
find $\Gamma_{\rm cusp}(g^2)$ 
from $W_4$ directly, without introducing a regularization, 
by computing
its second derivative \re{dd}.

We would like to emphasize that  relations \re{Z} -- \re{dd} follow from the conformal Ward identities and should hold in the presence of instantons. We shall verify this property below.

We recall that the instanton correction to $W_4$ should match the leading correction to the correlation function
$\mathcal G(u,v)$  for $u,v\to 0$. Taking into account \re{G8}, we expect that the instanton correction to the
cusp anomalous dimension should scale as  $\Gamma_{\rm cusp}(g^2)=O(g^8 \e^{2\pi i \tau})$.

\subsection{Semiclassical approximation}    
 
To compute instanton corrections to the light-like Wilson loop, we have to define $W_4$ in Euclidean signature.
This can be achieved by allowing the cusp points $x_i^\mu$ to take complex values, such that $x_{i,i+1}^2=0$. Having determined 
$W_4$ as a function of $x_{13}^2$ and $x_{24}^2$, we shall  continue it to Minkowski signature. 

To find $W_4$ in the semiclassical approximation, we have to evaluate the Wilson loop \re{W4-A}
in the instanton background and, then, integrate it over the collective coordinates with the measure \re{measure}
\begin{align}\label{W4-inst}
W_{4,\rm inst} = \int d\mu_{\rm phys} \e^{-S_{\rm phys}} W_{A}\,.
\end{align} 
We recall that $W_{A}$ is defined in the adjoint representation of the $SU(2)$. 
It proves convenient to generalize \re{W4-A} and define
the Wilson loop in an arbitrary $SU(2)$ representation $R$ 
\begin{align}\label{W_R}
W_{R} = \tr_R\left[E(x_1,x_2) E(x_2,x_3) E(x_3,x_4) E(x_4,x_1)  \right]\,,
\end{align}
where $E(x_i,x_{i+1})$ is a light-line Wilson line stretched between the points $x_i$ and $x_{i+1}$ 
\begin{align}\label{E}
E(x_i,x_{i+1}) = P\exp\lr{-ig \int_0^1 dt\, x^\mu_{i,i+1} A_\mu(x_i - t x_{i,i+1})}\,.
\end{align}
Here the gauge field $A_\mu(x) = A_\mu^a(x) T^a$  is integrated
along the light-cone segment $x(t) = x_i - t x_{i,i+1}$  and $T^a$ are the $SU(2)$ generators in the representation $R$.
Notice that for zero value of the coupling constant, the Wilson loop \re{W_R} is equal to
the dimension of the representation $W_{R}=d_R$. 

In special cases of the fundamental $(F)$ and adjoint $(A)$ representations of the $SU(2)$,  the generators  
are related to Pauli matrices, $T^a=\sigma^a/2$, and completely antisymmetric tensor,  $(T^a)^{bc} = i\epsilon^{abc}$, respectively.
The corresponding Wilson loops, $W_{F}$ and $W_{A}$, satisfy the fusion relation
\begin{align}\label{WF-WA}
W_{F} \widebar W_{F}  = 1 + W_{A} \,,
\end{align}
where $\widebar W_{F}$ is complex conjugated to $W_F$.  

Applying \re{W4-inst}, we have to replace the gauge field in \re{W_R} with its expression in the instanton background, 
$A(x)= g^{-1} A_{\rm inst}(x)$ (see Eqs.~\re{dec} and \re{A-exp}). As follows from  \re{W_R} and \re{E}, the resulting expression for $W_R$ 
does not depend on the coupling constant. It depends however on $16$ fermion modes of the instanton, $\xi_\alpha^A$ and 
$\bar\eta_{\dot\alpha}^A$. This dependence has the following general form
\begin{align}\label{WF-dec}
W_F = W^{(0)} +  W^{(4)} +  W^{(8)}+  W^{(12)}+  W^{(16)}\,,
\end{align}
where $W^{(4n)}$ denotes a homogenous $SU(4)$ invariant polynomial in $\xi_\alpha^A$ and $\bar\eta_{\dot\alpha}^A$ of degree $4n$.
The Wilson loop in the adjoint representation has similar form. The relation \re{WF-WA} allows us to express $W_A$ in terms of 
$W^{(4n)}$. 

In order to compute the Wilson loop \re{W4-inst} we only need the top component $W_R^{(16)}$ 
containing $16$ fermion modes. The remaining components give vanishing contribution upon integration over fermion modes in \re{W4-inst}.
For the Wilson loop in the fundamental representation, the top component $W_F^{(16)}$ is given by $W^{(16)}$. For the Wilson loop in the adjoint representation we get from \re{WF-WA}
and \re{WF-dec}
\begin{align}\label{WA-dec}
W_A^{(16)}  = W^{(0)} \widebar W^{(16)} + W^{(4)} \widebar W^{(12)} + \frac12 W^{(8)} \widebar W^{(8)} + \text{c.c.}  \,.
\end{align}     
Since $W_R^{(16)}$ contains $16$ fermion modes, it has the following form
\begin{align}\label{F_R}
W_R^{(16)} = f_R(x_i; x_0,\rho)\, \xi^8 \,\bar\eta^8\,,
\end{align}
where $\xi^8 = \prod_{\alpha, A} \xi_\alpha^A$ and similar for $\bar\eta$. The scalar function $f_R(x_i; x_0,\rho)$ depends on 
four points $x_i$ (with $x_{i,i+1}^2=0$) and on the bosonic collective coordinates $x_0$ and $\rho$. It also depends on the
representation $R$ of the $SU(2)$ gauge group.

Substituting \re{F_R} into \re{W4-inst}
and taking into account \re{measure}, we obtain the following expression for the instanton correction to $W_4$ in the semiclassical 
approximation \footnote{Notice that $W_{4,\rm inst}$ depends on the choice of the $SU(2)$  representation $R$.}
\begin{align}\label{W-form}
W_{4,\rm inst} = {g^8 \over 2^{34} \pi^{10}} \e^{2\pi i\tau} \int d^4 x_0 \int{d\rho\over \rho^5} f_R(x_i; x_0,\rho)\,.
\end{align}
The dependence of this expression on the coupling constant matches \re{G8}. We expect that the instanton effects
should modify the light-like cusp anomalous dimension. For this to happen, the integral in \re{W-form} has to develop UV divergences.
Indeed, as we show below, instantons of small size ($\rho\to 0$) located in the vicinity of the cusp points ($x_0\to x_i$) provide a
divergent contribution to \re{W-form}.
      
\subsection{Cusp anomalous dimension}     
     
The light-like Wilson loop is invariant under conformal transformations at the classical level. At the quantum level, its conformal symmetry
is broken by cusp singularities. In application to \re{W-form} this implies that if the integral in \re{W-form} were well-defined,
$W_{4,\rm inst}$ should be conformally invariant. The conformal transformations act nontrivially on the bosonic moduli $x_0$ and $\rho$
leaving the integration measure in \re{W-form} invariant. Therefore, invariance of $W_{4,\rm inst}$ under these transformations 
translates into conformal invariance of the function $f_R(x_i; x_0,\rho)$. To regularize cusp singularities of \re{W-form} we can 
modify the integration measure as 
\begin{align}\label{dx0}
\int d^4 x_0 \ \to \ \mu^{-2\epsilon}\! \int d^{4-2\epsilon} x_0\,,
\end{align}
leaving the function $f_R(x_i; x_0,\rho)$ intact.~%
\footnote{There are of course different ways to regularize \re{W-form}. The relation \re{dd} ensures that the resulting expression for the cusp anomalous dimension
is independent on the regularization procedure. This does not apply however to the collinear anomalous dimension.}

The conformal symmetry dictates that the function $f_R(x_i; x_0,\rho)$ can depend on the bosonic moduli, $\rho$ and $x_0$, and four 
points $x_i$  through conformal invariants only. The latter have the following form 
\footnote{To check  conformal properties of $I_{ij}$ we can employ inversions defined in \re{inv}.  It is straightforward to verify that \re{Inv} is invariant under these transformations.}
\begin{align}\label{Inv}
I_{ij} = {x_{ij}^2\rho^2\over (x_{i0}^2+\rho^2)(x_{j0}^2+\rho^2)}\,.
\end{align}    
 We recall that the points $x_i$ define the
vertices of light-like rectangle and satisfy $x_{i,i+1}^2=0$. As a consequence, $I_{i,i+1}=0$ and we are left with only two nonvanishing
invariants, $f_R = f_R(I_{13},I_{24})$. 

Since $f_R$ is obtained from the Wilson loop \re{F_R} evaluated in background of instanton field, it is an intrinsically classical quantity. 
We therefore expect it to be a rational function of  $I_{13}$ and $I_{24}$. In addition, $f_R$ should vanish for $I_{13}=0$
or $I_{24}=0$. The reason for this is that for $x_1\to x_3$ (or $x_2\to x_4$) the 
rectangular contour in \re{W_R} collapses into a closed backtracking path. The Wilson lines in \re{W_R} cancel against each other
for such path leading to $W_R=d_R$ or equivalently $f_R=0$. 
These properties suggest to look for $f_R(I_{13},I_{24})$ in the form
\begin{align}\notag\label{f-gen}
f_R (I_{13},I_{24}) {}&= \sum_{\ell_1,\ell_2\ge 1} f_{\ell_1 \ell_2}\, (I_{13})^{\ell_1}\, (I_{24})^{\ell_2}
\\
{}&= \sum_{\ell_1,\ell_2\ge 1}   { f_{\ell_1 \ell_2}\rho^{2(\ell_1+\ell_2)} (x_{13}^2)^{\ell_1}(x_{24}^2)^{\ell_2}\over 
[(x_{10}^2+\rho^2)(x_{30}^2+\rho^2)]^{\ell_1}[(x_{20}^2+\rho^2)(x_{40}^2+\rho^2)]^{\ell_2}} \,,
\end{align}            
where expansion coefficients are symmetric $f_{\ell_1\ell_2}=f_{\ell_2\ell_1}$ due to the cyclic symmetry of \re{W_R}. Moreover, as we show in Appendix~\ref{app:gauge}, $f_R (I_{13},I_{24})$ is actually a polynomial in both variables, so that the sum in \re{f-gen} contains a finite number of terms.
     
Replacing the function $f_R$ in \re{W-form} with its general expression  \re{f-gen}, we find that the integrals over $x_0$ and $\rho$ can be 
expressed in terms of $D-$functions defined in Appendix~\ref{app:D-fun}. These functions are finite for generic $x_{ij}^2\neq 0$ but
develop logarithmic divergences for $x_{i,i+1}^2=0$. A close examination shows that divergences arise from integration over 
$\rho\to 0$ and $x_0\to x_i$ and have a clear UV origin. These are the cusp divergences that were mentioned at the end of the
previous subsection. Regularizing divergences according to \re{dx0}, we obtain
 the following expression for the instanton correction to the light-like Wilson loop \re{W-form} 
\begin{align}\label{W-sum}
W_{4,\rm inst} = {g^8 \over 2^{34} \pi^{10}} \e^{2\pi i\tau}\sum_{\ell_1,\ell_2} f_{\ell_1 \ell_2}\,\bar D^{(\epsilon)}_{\ell_1\ell_2}\,.
\end{align}
Here we introduced notation for the regularized integral
\begin{align}\label{d-eps}
\bar D^{(\epsilon)}_{\ell_1\ell_2}=\mu^{-2\epsilon}\int d^{4-2\epsilon} x_0 \int{d\rho\over \rho^5}  { \rho^{2(\ell_1+\ell_2)} (x_{13}^2)^{\ell_1}(x_{24}^2)^{\ell_2}\over 
[(x_{10}^2+\rho^2)(x_{30}^2+\rho^2)]^{\ell_1}[(x_{20}^2+\rho^2)(x_{40}^2+\rho^2)]^{\ell_2}}\,,
\end{align}  
evaluated for $x_{i,i+1}^2=0$.  This integral is well-defined for $\epsilon<0$ and the cusp divergences appear as poles in $\epsilon$. The details of calculation can be found in Appendix~\ref{app:D-fun}.

Substituting the resulting expression  for $\bar D^{(\epsilon)}_{\ell_1\ell_2}$ (see \re{D-eps-res} in
Appendix~\ref{app:D-fun})  into \re{W-sum} we find that $W_{4,\rm inst}$ takes a remarkable simple form
\begin{align}\notag\label{W-col}
W_{4,\rm inst}/d_R = {}&-   \Gamma_{\rm cusp}(g^2) \bigg[{1\over \epsilon^2}  (\mu^2 x_{13}^2)^{-\epsilon} +{1\over \epsilon^2}  (\mu^2 x_{24}^2)^{-\epsilon}  -\frac12\ln^2(x_{13}^2/x_{24}^2) \bigg] 
\\
  {}&- \Gamma_{\rm col}(g^2)\bigg[{1\over \epsilon}  (\mu^2 x_{13}^2)^{-\epsilon} +{1\over \epsilon}  (\mu^2 x_{24}^2)^{-\epsilon}   \bigg] - \Gamma_{\rm h}(g^2) \,,
\end{align}
where $d_R$ is the dimension of the $SU(2)$ representation in which the Wilson loop is
defined.  Here we denoted the residue at the double pole as 
$\Gamma_{\rm cusp}(g^2)$ anticipating that the same quantity defines the instanton correction to the cusp anomalous dimension. 
It is given by the following expression
\begin{align}\label{cusp-f}
 \Gamma_{\rm cusp}(g^2) = - {g^8\over 2^{34} \pi^{8}d_R} \e^{2\pi i\tau}    \sum_{\ell_1,\ell_2\ge 1} f_{\ell_1 \ell_2}   {\Gamma\lr{ \ell_1+\ell_2-2} \over \Gamma(\ell_1)\Gamma(\ell_2)} \,.
\end{align} 
We recall that $f_{\ell_1 \ell_2}$ are coefficients of the expansion of the top component of the Wilson
loop \re{F_R} in powers of the conformal invariants \re{f-gen}. 
Notice that for $\ell_1=\ell_2=1$ the $\Gamma-$function in \re{cusp-f} develops a pole. For the sum in \re{cusp-f} 
to be finite the corresponding coefficient $f_{11}$ has to vanish.

The residue at the simple pole and the constant term in \re{W-col}, $\Gamma_{\rm col}(g^2)$ and $\Gamma_{\rm h}(g^2)$, respectively, are given by  expressions similar to \re{cusp-f}. However, in distinction from $\Gamma_{\rm cusp}(g^2)$ they depend on the choice of the regularization in \re{d-eps}. That is why we do not present their expressions.

Let us now compare \re{W-col} with the expected properties of light-like Wilson loop. We combine \re{W-col} with the Born level
contribution to the Wilson loop, $W_4=d_R + W_{4,\rm inst}$,  and require that  $W_4$ has to satisfy \re{Z} and \re{dd}.
This leads to the following relations 
\begin{align}\label{dd1}
{}& \frac12 \lr{{\partial \over \partial  \ln \mu^2}}^2\lr{W_{4,\rm inst}/d_R}
=
{\partial \over \partial  \ln x_{13}^2}{\partial \over \partial \ln x_{24}^2} \lr{W_{4,\rm inst}/d_R} = -  \Gamma_{\rm cusp}(g^2) \,.
\end{align}
It is easy to check that \re{W-col} verifies these relations. In this way, we find that \re{cusp-f} defines indeed the leading instanton correction to the cusp anomalous dimension.
  
In addition to \re{W-col}, the light-like Wilson loop also receives perturbative corrections that run in powers of $g^2$. To lowest order in $g^2$, these corrections have exactly the same form \re{W-col} although expressions for the anomalous
dimensions are different \cite{Drummond:2007aua}. The reason for such universality can be understood as follows.
As  explained in Section~\ref{sect:conf}, the conformal symmetry fixes the dependence of the light-like rectangular Wilson loop
on kinematical invariants. In particular, it allows us to determine the finite part of the Wilson loop in terms of the cusp anomalous 
dimension, Eq.~\re{F4}. The fact that the instanton corrections \re{W-col} verify \re{Z} and \re{dd} implies that the conformal
Ward identities found in \cite{Drummond:2007au} hold nonperturbatively, in the presence of instanton effects.  
  
In the next section, we apply \re{cusp-f}
to compute the leading instanton contribution to the cusp anomalous dimension  in the fundamental and adjoint representations
of the $SU(2)$. 

\section{Instanton contribution to the cusp anomalous dimension}       

According to \re{W-sum}, the instanton corrections to the light-like Wilson loop are determined by the
coefficients $f_{\ell_1 \ell_2}$. To find them, we have to identify the top component of the Wilson loop \re{F_R}
containing $16$ fermion modes and, then, expand the corresponding function $f_R$ in powers of the conformal invariants \re{f-gen}.

The top component of the Wilson loop $W_R^{(16)}$ depends on the choice of the representation $R$. 
Making use of the relation \re{WA-dec} (and its generalization for higher spin representations of the $SU(2)$) we can express
$W_R^{(16)}$ in terms of various components $W^{(4n)}$ of the Wilson loop in the fundamental 
representation \re{WF-dec}.  
 
\subsection{Wilson loop in the fundamental representation} 
 
To compute the Wilson loop in the fundamental representation of the $SU(2)$, we have to replace the gauge field in \re{W_R} and 
\re{E} with its expression \re{A-exp} in the instanton background, $A^\mu(x) =g^{-1} A^{\mu, a}_{\rm inst}(x) \,\sigma^a/2$. 

The resulting expression for the Wilson line \re{E} depends on $16$ fermion modes and admits an expansion similar to \re{WF-dec}
\begin{align}\label{E-dec}
E(x_i,x_{i+1}) = P \e^{i \int_{x_i}^{x_{i+1}} dx_\mu  A^\mu_{\rm inst}(x)} =  E^{(0)} +  E^{(4)} +  E^{(8)} + E^{(12)} + E^{(16)}   \,.
\end{align}
In distinction from \re{WF-dec}, each term on the right-hand side is gauge dependent. 
The explicit expressions for the first three terms on the right-hand side of \re{E-dec} are
\begin{align}\notag\label{Es}
E^{(0)}(x_i,x_{i+1}) {}& =  P \exp\lr{i \int_0^1 dt\, \dot x^\mu (t) A_\mu^{(0)}(x(t))}\,,
\\\notag
E^{(4)}(x_i,x_{i+1}) {}& = i\int_0^1 dt \,\dot x^\mu (t) E^{(0)}(x_i,x(t))  A_\mu^{(4)}(x(t)) E^{(0)}(x(t),x_{i+1}) \,,
\\\notag
 E^{(8)}(x_i,x_{i+1}) {}& = i\int_0^1 dt \,\dot x^\mu (t) E^{(0)}(x_i,x(t))  
 \\[1.5mm]
 {}& \times \left[ A_\mu^{(8)}(x(t)) E^{(0)}(x(t),x_{i+1}) + A_\mu^{(4)}(x(t)) E^{(4)}(x(t),x_{i+1})\right],
\end{align}
where $x(t) = (1-t) x_i + t x_{i+1}$ parameterizes the light-like segment $[x_i,x_{i+1}]$ and $\dot x(t) = \partial_t x(t) = -x_{i,i+1}$. Expressions for the remaining components
of \re{E-dec} are more involved. Going through their calculation we find (see Appendix~\ref{app:conformal}) that they are proportional to the square of a fermion mode and, therefore, have to vanish
\begin{align}\label{E12}
 E^{(12)} (x_i,x_{i+1}) = E^{(16)}(x_i,x_{i+1}) = 0\,.
\end{align}
Thus, the light-like Wilson line \re{E-dec} has at most $8$ fermion modes.

Substituting \re{E-dec} into \re{W_R} we obtain
that the Wilson loop \re{WF-dec} is given by a linear combination of terms of the form
\begin{align}\label{W-ind} 
W^{(k_1,k_2,k_3,k_4)} {}&= \tr\left[E^{(k_1)}(x_1,x_2) E^{(k_2)}(x_2,x_3)E^{(k_3)}(x_3,x_4)E^{(k_4)}(x_4,x_1)\right],
\end{align}
where integers $k_1,\dots,k_4$ count the number of fermion modes. 
The explicit expressions for different components of \re{WF-dec} are
\begin{align}\notag\label{Ws}
{}& W^{(0)} = W^{(0,0,0,0)},
\\[1.5mm]\notag
{}& W^{(4)} = W^{(4,0,0,0)} + \text{cyclic},
\\
{}& W^{(8)} = W^{(8,0,0,0)} + W^{(4,4,0,0)} + \frac12 W^{(4,0,4,0)} + \text{cyclic},
\\ \notag
{}& W^{(12)} = W^{(8,4,0,0)} + W^{(8,0,4,0)} + W^{(8,0,0,4)}+ W^{(4,4,4,0)} + \text{cyclic},
\\\notag
{}& W^{(16)} = W^{(8,4,4,0)} + W^{(8,4,0,4)} + W^{(8,0,4,4)} + W^{(8,8,0,0)} + \frac12 W^{(8,0,8,0)} + \frac14 W^{(4,4,4,4)} + \text{cyclic},
\end{align}
where `cyclic' denotes the additional terms that ensure the invariance of Wilson loop under the cyclic shift of the cusp points $(x_1,\dots,x_4)$. The additional 
rational factors are inserted to avoid a double counting.

By definition, $W^{(4n)}$ is a homogenous polynomial in fermion modes $\xi_\alpha^A$ and $\bar\eta_{\dot\alpha}^A$ 
of degree $4n$. It depends in addition on four cusp points $x_i$ and bosonic moduli $x_0$ and $\rho$. 
As a function of these variables, it should be invariant under conformal transformations including inversions 
(see \re{inv}, \re{inv-fer} and \re{inv-W4} in Appendix~\ref{app:conformal}). We shall use this property below to simplify the calculation. 
  
\subsection{Leading term} 
 

It is convenient to switch from vector to spinor notations and convert $A_\mu(x)$ into $2\times 2$ matrix $A_{\alpha\dot\alpha}(x)$
by contracting its Lorentz index with four-dimensional vector of Pauli matrices $\sigma^\mu = (1,i \boldsymbol{\sigma})$
\begin{align}
(A_{\alpha\dot\alpha})_i{}^j = i A^a_\mu(x)(\sigma^a/2)_i{}^j (\sigma^\mu)_{\alpha\dot\alpha}\,.
\end{align} 
This field carries two spinor indices ($\alpha,\dot\alpha=1,2$) and two 
$SU(2)$ indices ($i,j=1,2$). In addition, we define the gauge field with lower $SU(2)$ indices,
$A_{\alpha\dot\alpha,ik}=\epsilon_{kj}(A_{\alpha\dot\alpha})_i{}^j$, 
it is symmetric with respect to indices $i$ and $k$.
 
The BPST instanton $A^{(0)}$ takes the following form in the spinor notations
\begin{align}\label{BPST}
A^{(0)}_{\alpha\dot\alpha,ij}(x) = {\epsilon_{i\alpha}(x-x_0)_{j\dot\alpha} + \epsilon_{j\alpha}(x-x_0)_{i\dot\alpha} \over (x-x_0)^2+\rho^2}\,,
\end{align}       
where $x_{\alpha\dot\alpha} = x_\mu (\sigma^\mu)_{\alpha\dot\alpha}$ for arbitrary four-dimensional vector $x_\mu$. 
It intertwines the $SU(2)$ gauge group and the chiral half of the Lorentz group $SO(4) = SU(2)\times SU(2)$.

To evaluate the leading term $E^{(0)}(x_1,x_2)$ defined in \re{Es} we make use of the identity
\begin{align} \label{sam}\notag
i\int  dt \, \dot x^\mu (t) A^{(0)}_\mu (x(t)) {}&=  - \frac12 \int  dt \, \dot x^{\dot\alpha\alpha} (t) A^{(0)}_{\alpha\dot\alpha} (x(t))
\\
{}&
= \frac12 \int {dt \,  [x_{10},x_{20}] \over (1-t)x_{10}^2 + t x_{20}^2 +\rho^2}\,,  
\end{align}
where $x(t)=(1-t)x_1+ t x_2$ and the $SU(2)$ indices are suppressed for simplicity. Here in the second relation we substituted \re{BPST}
and contracted Lorentz indices using the conventions described in Appendix~\ref{app:conv}. Since the $SU(2)$ matrix part of the integral \re{sam} does not depend on the integration variable, the path-ordered exponential in \re{Es} reduces to the conventional exponential. Its calculation yields 
\begin{align}\label{E0}
E^{(0)}(x_1,x_2) = {\rho^2+x_{10}x_{20}\over [(x_{10}^2+\rho^2)(x_{20}^2+\rho^2)]^{1/2}}\,.
\end{align}
Here  the first term in the numerator is proportional to the $SU(2)$ identity matrix $\delta_i^j$ and the second one involves the matrix $(x_{10}x_{20})_i{}^j = (x_{10})_{i\dot\alpha} (x_{20})^{\dot\alpha j}$.

We can now use \re{E0} to compute the lowest component $W^{(0)}$ of the light-like Wilson loop \re{Ws}.
Going through the calculation we find
\begin{align}
W^{(0)} =2 -  {\rho^4 x_{13}^2 x_{24}^2 \over (\rho^2+x_{01}^2)(\rho^2+x_{02}^2)(\rho^2+x_{03}^2)(\rho^2+x_{04}^2) } = 2- I_{13} I_{24}\,,
\end{align}
where $I_{ij}$ are conformal invariants defined in \re{Inv}. Their appearance is not surprising since $W^{(0)}$ should be
invariant under the conformal transformations.

\subsection{Conformal gauge}\label{sect:gauge}

The calculation of the remaining components in \re{Es} is more involved since the gauge
fields $A^{(4)}$ and $A^{(8)}$ are given by complicated expressions (see Appendix~\ref{app:gauge}). It can be significantly simplified by 
making use of the conformal symmetry. 

We can exploit this symmetry to choose the cusp points $x_i$ to satisfy
the following additional conditions
\begin{align}\label{c-gauge}
x_{10}^2=x_{20}^2=x_{30}^2=x_{40}^2=0\,.
\end{align}
The advantage of this gauge is that any point of the light-like rectangle $x(t) = (1-t) x_i + t x_{i+1}$ 
becomes null separated from the center of instanton $x_0$
\begin{align}
(x(t)-x_0)^2= (1-t) x_{i0}^2 + t x_{i+1,0}^2 = 0 \,,
\end{align}
where we took into account that $x_{i,i+1}^2=0$.

The calculation of the Wilson line \re{Es} in the conformal gauge \re{c-gauge} is described in 
details in Appendix~\ref{app:gauge}. 
For the component $E^{(4)}(x_1,x_2)$ defined in \re{Es} we find 
\begin{align}\notag\label{E4}
E^{(4)}(x_1,x_2) 
{}& = {2\over 3\rho^2}   
\epsilon_{ABCD}  \bra{\zeta_1^{A}} E_{12}\ket{\zeta_2^{B}}
\\
{}& \times
\Big(\ket{\zeta_1^C} \bra{\zeta_2^D} \, + E_{12}\ket{\zeta_2^C} \bra{\zeta_1^D}E_{12}+ 2 \ket{\zeta_1^C} \bra{\zeta_1^D}E_{12} +2 E_{12} \ket{\zeta_2^C} \bra{\zeta_2^D}  \Big),
\end{align}
where $E_{12} \equiv E^{(0)}(x_1,x_2)$ and $\zeta_i$ is a linear combination of fermion modes, 
$\zeta_i = \xi + x_i \bar\eta$. The expression in the second line of \re{E4} contains the sum of four $SU(2)$ tensors,
each given by the direct product of two pairs of vectors, $\ket{\zeta_1}$, $E_{12}\ket{\zeta_2}$ and
$\bra{\zeta_2}$, $\bra{\zeta_1}E_{12}$. 

As we show in Appendix~\ref{app:conformal}, the general form of \re{E4} and
of the remaining components of the Wilson line   is fixed by the conformal symmetry. 
 In particular, $E^{(8)}(x_1,x_2)$ has the following form in the conformal gauge \re{c-gauge}
\begin{align}\notag\label{E8}
E^{(8)}(x_1,x_2)  = {1\over \rho^{4}}  
\Big({}& P_{CD}\ket{\zeta_1^C} \bra{\zeta_2^D} \,+ \, Q_{CD}  E_{12}\ket{\zeta_2^C} \bra{\zeta_1^D}E_{12}
\\
{}& + R_{CD} \ket{\zeta_1^C} \bra{\zeta_1^D}E_{12} + S_{CD}  E_{12} \ket{\zeta_2^C} \bra{\zeta_2^D}  \Big),
\end{align}
where
$P_{CD}, Q_{CD}, R_{CD}, S_{CD}$ are homogenous polynomials in $\zeta_1$ and $\zeta_2$ of degree $6$.
In virtue of conformal symmetry, the dependence on $\zeta_1$ and $\zeta_2$ can only enter through the
following three combinations
\begin{align}\label{zeta-zeta}
 \vev{\zeta_1^{A} \zeta_1^B}\,,\qquad  \qquad 
\vev{\zeta_2^{A} \zeta_2^B}\,,\qqqquad 
 \bra{\zeta_1^{A}} E_{12}\ket{\zeta_2^{B}}\,,
\end{align}
where we used notations for $  \vev{\zeta_1^{A} \zeta_1^B} \equiv \zeta_1^{\alpha A} \zeta_{1,\alpha}^B$ and
$\bra{\zeta_1^{A}} E_{12}\ket{\zeta_2^{B}}\equiv \zeta_1^{i A} (E_{12})_i{}^j \zeta_{2,j}^B$.
The power of $\rho$  in \re{E8} is fixed by the condition for $E^{(8)}(x_1,x_2)$ to be dimensionless.
The explicit expressions for the polynomials $P_{CD},\dots,S_{CD}$ are cumbersome, to save space we do not present them here. 
 
\subsection{Results}\label{sect:res}
   
We can use the expressions for the Wilson line obtained in the previous subsection to compute different components of the 
Wilson loop  in the fundamental representation of the $SU(2)$, Eqs.~\re{W-ind} and \re{Ws}. We recall that in order to find
the instanton correction to the Wilson loop we only need the top component $W^{(16)}$. It has the general form \re{F_R} and \re{f-gen} and
is specified by the set of coefficients $f_{\ell_1\ell_2}$.
 
The calculation of \re{W-ind} and \re{Ws} is rather lengthy and can be performed with a help of {\sl Mathematica}.     
This yields the following result for the coefficients  $f_{\ell_1\ell_2}$  (with $\ell_1,\ell_2\ge 1$)
\begin{align}\label{f-F}
f^{(F)}_{\ell_1\ell_2}=2^{16}\times \left(
\begin{array}{ccccc}
 0 & 0 & -288 & 0 & 0 \\
 0 & -144 & 0 & 288 & 0 \\
 -288 & 0 & 192 & 0 & -36 \\
 0 & 288 & 0 & -66 & -12 \\
 0 & 0 & -36 & -12 & -1 \\
\end{array}
\right)\,,
\end{align}
where we inserted the superscript to indicate that these coefficients define $W_F^{(16)}$ in the fundamental representation.

For the Wilson loop in the adjoint representation of the $SU(2)$ the calculation of \re{WA-dec} leads to
\begin{align}\label{f-A}
f^{(A)}_{\ell_1\ell_2}=2^{20}\times \left(
\begin{array}{cccccc}
   0 & 0 & -72 & 0 & 0 & 0 \\
0 & 144 & -360 & 312 & 0 & 0 \\
 -72 & -360 & 408 & 120 & -144 & 0 \\
 0 & 312 & 120 & -150 & -48 & 16 \\
 0 & 0 & -144 & -48 & 24 & 8 \\
  0 & 0 & 0 & 16 & 8 & 1 \\
\end{array}
\right)\,.
\end{align}
We observe that $f_{11}$ vanishes for both matrices. This ensures a finiteness of the sum in \re{cusp-f}.
       
Finally, we substitute \re{f-F} and \re{f-A} into \re{cusp-f}, replace $d_F=2$, $d_A=3$ and obtain the one-instanton correction to the cusp anomalous dimension
in the fundamental and adjoint representations of the $SU(2)$
\begin{align}\notag\label{cusps}
{}& \Gamma^{(F)}_{\rm cusp} = {387\over 8192} \lr{g^2\over 4\pi^2}^4\lr{\e^{2\pi i\tau} +\e^{-2\pi i\bar\tau} }\,,
\\
{}& \Gamma^{(A)}_{\rm cusp} = -{4\over 15} \lr{g^2\over 4\pi^2}^4\lr{\e^{2\pi i\tau} +\e^{-2\pi i\bar\tau} }\,,
\end{align}
where we  added the contribution of anti-instanton.  
The following comments are in order.

The relations \re{cusps} define nonperturbative corrections to the cusp anomalous dimension for two different representations of the
$SU(2)$ group. Perturbative corrections to $ \Gamma_{\rm cusp}(g^2)$ are known to verify the so-called Casimir scaling up to order
 $O(g^6)$. Namely, perturbative contribution to  $ \Gamma_{\rm cusp}(g^2)$ depends
 on the representation $R$ through the 
quadratic Casimir only, $  \Gamma_{\rm cusp}(g^2)\sim C_R$.  This property is violated however at order $O(g^8)$ due to the appearance of higher Casimirs 
\cite{Frenkel:1984pz}. We can easily check using \re{cusps} that the instanton corrections do not verify the Casimir scaling.
If this property were true, the ratio of two expressions in \re{cusps} would be equal to the ratio of the quadratic Casimir operators  $C_F/C_A$
with $C_F=3/4$ and $C_A=2$ in the fundamental and adjoint representations of the $SU(2)$, respectively. Obviously, the
expressions \re{cusps} do not have this property, not to mention that the instanton corrections have an opposite sign for the two
representations.

The relations \re{cusps} describe the leading instanton correction to the cusp anomalous dimension for the $SU(2)$ gauge group. 
Following \cite{Dorey:1998xe}, we can generalize them to the $SU(N)$  gauge group. In this case, the instantons have the additional
$4N-8$ bosonic modes describing the embedding of the $SU(2)$ instanton into the $SU(N)$ and $8N-16$ `nonexact' fermionic modes. Their contribution amounts to multiplying \re{cusps} by the factor of $(2N-2)!/[2^{2N-3}(N-1)!(N-2)!]$. For large $N$, the
relations \re{cusps} can be also extended to the multi-instanton sector \cite{Dorey:1999pd}. In this limit, the integral over the moduli space of instantons
is dominated by the saddle point in which all instantons are at the same position $x_0$, have the same size $\rho$ and lie 
in commuting $SU(2)$ blocks inside the $SU(N)$. Up to overall $O(g^8)$ factor, the resulting expressions are similar to 
those given in \cite{Dorey:1999pd}.
        
\section{Concluding remarks}          

In this paper, we have studied instanton corrections to the four-point correlation correlation function of half-BPS operators in $\mathcal N=4$ SYM 
in the light-cone limit when operators become null separated in a sequential manner. Perturbative corrections to the correlation function 
in this limit are enhanced by logarithms of vanishing cross ratios and can be summed to all orders in the coupling. Previous studies revealed 
that in the semiclassical approximation the instanton corrections are suppressed  in the light-cone limit  by powers of the cross ratios.
It is natural to ask whether this result is an artefact of the approximation or an intrinsic feature of instantons. 

To answer this question 
we exploited the relation between the leading asymptotic behaviour of the correlation function and light-like rectangular Wilson loop. 
Analysing this relation we found an interesting interplay between semiclassical and quantum instanton corrections. 
\footnote{Similar phenomenon has been previously observed for the Konishi
operator \cite{Alday:2016tll}.}
Namely, 
having computed the light-like Wilson loop in the semiclassical approximation, we were able to identify the leading instanton contribution
to the correlation function in the light-cone limit. In the conventional approach, the same correction would correspond to  
taking into account the contribution to the correlation function of quantum fluctuations of instantons to forth order in perturbation theory.
 
We also demonstrated that the light-like rectangular Wilson loop satisfies conformal Ward identities and identified the leading 
instanton correction to the cusp anomalous dimension. Making use of this result, we can determine the leading instanton contribution to anomalous dimension of twist-two operators with large spin \re{gamma-S} and answer the question of how instantons modify the light-cone asymptotic behaviour of the four-point correlation
function \re{G-lc}.  

At weak coupling, $\mathcal G(u,v)$ receives both perturbative and instanton corrections enhanced by 
powers of $\ln u$ and $\ln v$. They arise due to logarithmic scaling \re{gamma-S} of the anomalous dimension of  the twist-two 
operators with large spin, $S\sim u^{-1/2}$ or $S\sim v^{-1/2}$, exchanged in different OPE channels. Following \cite{Alday:2010zy,Alday:2013cwa}, 
such logarithmically enhanced corrections can be resummed leading to 
\begin{align}\label{G-res}
\mathcal G(u,v) \sim \exp\lr{-{\hat u \,\hat v\over 2\Gamma_{\rm cusp}(g^2) }} \times \e^{-2 \Gamma_{\rm cusp}(g^2) \partial_{\hat u}  \partial_{\hat u} } 
\left[\e^{\gamma_{_{\rm E}} (\hat u+\hat v)/2}\Gamma\lr{1-\frac 12 \hat u} \Gamma\lr{1-\frac 12 \hat v} \right]^2,
\end{align}
where 
$ \hat u = \gamma_{S=1/\sqrt{u}}$ and $\hat v = \gamma_{S=1/\sqrt{v}}$ are the anomalous dimensions \re{gamma-S} evaluated for the 
values of spins mentioned above, $\Gamma_{\rm cusp}(g^2)$ is given by the sum of perturbative and instanton contributions
and $\gamma_{_{\rm E}}$ is Euler's constant. The 
relation \re{G-res} has the expected factorized form \re{G-lc}, the first factor on the right-hand side of \re{G-res} comes from the
rectangular light-like Wilson loop $W_4$ whereas the second one from the jet function $J$. 

The relation \re{G-res} develops poles at even positive $\hat u$ and $\hat v$. These poles have a clear physical meaning and have
important consequences for the $S-$duality properties of the four-point correlation function.
We recall that \re{G-res} takes into account the contribution of twist-two operators only. For $\hat u=2$ (or $\hat v=2$) we encounter
a level crossing phenomenon \cite{Korchemsky:2015cyx} -- the twist-two operators acquire anomalous dimension $2$ and collide with the twist-four operators. The appearance of
spurious poles in \re{G-res} is a consequence of ignoring the contribution of the latter operators to \re{G-res}. To obtain 
a reliable prediction for $\mathcal G(u,v)$ in the vicinity of the pole we have to resolve the mixing of twist-two and 
twist-four operators and include the contribution of both to \re{G-res}. To avoid remaining poles of $\mathcal G(u,v)$ we have to 
take into account the mixing with operators of higher twist.

We can arrive at the same conclusion by examining properties of the both sides of \re{G-res} under the $S-$duality transformations. 
The relation between the correlation function and light-like Wilson loop $\mathcal G(u,v) \sim W_4$ cannot hold for an 
arbitrary coupling since the two quantities have different properties. Indeed, the $S-$duality maps
Wilson loop into 't Hooft loop while the four-point correlation function $\mathcal G(u,v)$ remains
invariant. To restore the $S-$duality of $\mathcal G(u,v)$, the higher twist contribution has to be added to \re{G-res}.
This problem deserves further investigation.

The above analysis can be extended to $n-$point correlation function of half-BPS operators $G_n$. At weak coupling, the leading asymptotics of $G_n$
in the light-cone limit is described by $n-$gon light-like Wilson loop $W_n$. It receives both perturbative and instanton corrections and
satisfies the conformal Ward identities \cite{Drummond:2007aua}. For $n=5$ the conformal symmetry uniquely fixes the form of $W_5$ in terms of the cusp anomalous dimension. For $n\ge 6$ it leaves a freedom of adding to $W_n$ a function of cross ratios, the so-called
remainder function. This function has been studied in planar $\mathcal N=4$ SYM where it was found to have 
a number of remarkable properties reflecting integrability of theory \cite{Basso:2013vsa}. It would be interesting 
to compute the leading instanton correction to the remainder function and to understand whether some of its symmetries
survive in the presence of nonperturbative effects.

The scattering amplitudes are known to be dual to the light-like Wilson loops in planar $\mathcal N=4$ SYM. One may wonder whether
the same relation holds for finite $N$ in the presence of instantons. The duality implies that infrared divergences of amplitudes should 
match ultraviolet (cusp) divergences of Wilson loops. We have shown in this paper that the instanton corrections to light-like Wilson loops
have the cusp divergences of the same double logarithmic form as in perturbation theory. This is not the case however for infrared divergences of the scattering
amplitudes. These divergences come from integration over instantons with large size $\rho$ and have a power-like dependence on the infrared
cutoff. The mismatch in the form of ultraviolet and infrared divergences points towards the breaking of the above mentioned duality between the instanton
contribution to scattering amplitudes and light-like Wilson loops. 
 
\section*{Acknowledgements}

We would like to thank Fernando Alday, Emery Sokatchev and Arkady Tseytlin for useful discussions.

\appendix

\section{Conventions}\label{app:conv}

Throughout the paper we use  Greek letters, $\alpha,\dot\alpha,\dots$ for Lorentz indices and Latin 
letters $i,j,\dots$ for the $SU(2)$  indices.

We use the Pauli matrices $\sigma^\mu_{\alpha\dot\alpha} = (1,i \boldsymbol{\sigma})$ to convert an arbitrary Euclidean four-vector $x_\mu=(x_1,x_2,x_3,x_4)$  into $2\times 2$ matrix
\begin{align}\label{x-mat}
x_{\alpha\dot\alpha} = x_\mu \sigma^\mu_{\alpha\dot\alpha} = \left[\begin{array}{cc} ix_3-x_4 & ix_1-x_2 \\ ix_1+x_2 & -ix_3-x_4\end{array}\right]\,.
\end{align} 
Its indices are raised and lowered with a help of an antisymmetric tensor  
\begin{align}\notag
{}& x^\alpha{}_{\dot\alpha} = \epsilon^{\alpha\beta} x_{\beta\dot\alpha}  \,, \qqqquad 
 x_\alpha{}^{\dot\alpha} =  x_{\alpha\dot\beta}\epsilon^{\dot\beta\dot\alpha}  \,,
 \qqqquad 
  x ^{\dot\alpha\alpha} = \epsilon^{\alpha\beta}x_{\beta\dot\beta}\epsilon^{\dot\beta\dot\alpha}  \,,
\end{align}
with $\epsilon_{\alpha\beta}\epsilon^{\gamma\beta} = \delta_\alpha^\gamma$ and $\epsilon_{12} = \epsilon^{12}=1$.
The product of matrices is defined as
\begin{align} \label{x1x2}
(x_1x_2)_{\alpha\beta}  = (x_1)_{\alpha\dot\alpha} (x_2)_\beta^{\dot\alpha} = (x_1)_{\alpha\dot\alpha}  (x_2)_{\beta\dot\beta} \epsilon^{\dot\beta\dot\alpha}  \,.
\end{align} 
For the scalar product of Euclidean vectors $(xy) = \sum_i x_i y_i$ we have
\begin{align}
(xy) = \frac12 x^\alpha_{\dot\alpha} y_\alpha^{\dot\alpha}=\frac12 x_{\alpha\dot\alpha} y_{\beta\dot\beta} \epsilon^{\alpha\beta}\epsilon^{\dot\alpha\dot\beta}\,.
\end{align}
Allowing vectors to have complex components, we can define Euclidean analog of light-like vectors $n_\mu$
satisfying $n^2=0$. In spinor notations $n_{\alpha\dot\alpha}$ factorizes into the product of 
commuting spinors $n_{\alpha\dot\alpha}=\ket{n} [n|$ leading to
\begin{align}
(n x n)_{\alpha\dot\alpha} = \ket{n} [n|x\ket{n} [n| = -2(xn) n_{\alpha\dot\alpha}\,,
\end{align}
where $ [n|x\ket{n} =n_{\alpha\dot\alpha} x^{\alpha\dot\alpha}=-2(xn)$.
Notice that in Minkowski signature the expression on the right-hand side
has an opposite sign. The reason for this is that the definition of the scalar product differs by
sign in Minkowski and Euclidean signatures, $(xy)_{_{\rm M}} =  - (xy)_{_{\rm E}}$ for $x_0 = i x_4$.

For the $SU(2)$ matrices $A_{i}{}^j $ we use similar conventions for raising and lowering indices
\begin{align}
  A_{i}{}^j = A_{ik}\epsilon^{kj} \,,\qqqquad A_{ij} = A_{i}{}^k  \epsilon_{jk}\,.
\end{align} 
The product of the $SU(2)$ matrices is defined as
\begin{align}\label{AB}
(A B)_{ik}  = A_i{}^j B_{jk}= A_{ij} \epsilon^{jl} B_{lk} 
\,,\qquad 
(A B C)_{in}  =A_i{}^j B_j{}^k C_{kn}
 \,.
\end{align}
In particular, the rectangular light-like Wilson loop $W_4$ is given by the product of four $SU(2)$ matrices
$E_{i,i+1}\equiv E(x_i,x_{i+1})$ defined as Wilson lines in the fundamental representation 
of the $SU(2)$ evaluated along the light-like segments $[x_i,x_{i+1}]$
\begin{align}\notag\label{W4-prod}
W_4 {}&=  (E_{12})_{i_1}{}^{i_2} (E_{23})_{i_2}{}^{i_3} (E_{34})_{i_3}{}^{i_4} (E_{41})_{i_4}{}^{i_1} 
\\[2mm]
{}&=  (E_{12})_{i_1k_1}\epsilon^{k_1i_2} (E_{23})_{i_2k_2}{\epsilon}^{k_2i_3} (E_{34})_{i_3k_3}{\epsilon}^{k_3i_4} (E_{41})_{i_4k_4}{\epsilon}^{k_4i_1} \,.
\end{align}
Computing instanton corrections we encounter matrices in the mixed representation, e.g. $(x_1 x_2)_{ij}$
with matrices $x_1$ and $x_2$ given by \re{x-mat}. By definition, they are given by expression 
like \re{x1x2} in which (chiral) Lorentz indices are identified with the $SU(2)$ indices
\begin{align}
(x_1 x_2)_{ij} \equiv (x_1)_{i\dot\alpha} (x_2)_j^{\dot\alpha} \,.
\end{align}
 
\section{Conformal properties}\label{app:conformal}

Since the equations of motion in $\mathcal N=4$ SYM are invariant under conformal transformations, these
transformations should map one classical solution into another one. 

To check conformal properties of instantons, we use inversions.  
They act on both space-time and collective coordinates of instantons
\begin{align}\label{inv}
I[ x_i^\mu] = {x_i^\mu\over x_i^2} \,,\qqqquad
I[ x_0^\mu] = {x_0^\mu\over x_0^2+\rho^2}\,,\qqqquad
I[ \rho] = {\rho\over x_0^2+\rho^2}\,.
\end{align}
For fermionic modes, the inversion is defined as
\begin{align}\label{inv-fer}
I [\xi_\alpha^A ]  = \bar\eta^A_{\dot\alpha}\,,\qqqquad I [\bar\eta^A_{\dot\alpha}] = \xi_\alpha^A\,.
\end{align}
The instanton fields depend on a linear combination of these modes
$\zeta=\xi + x\bar\eta$. As follows from \re{inv-fer}, it transforms covariantly under inversions  
\begin{align}\label{zeta-inv}
I[\zeta_\alpha^A(x)] = {x^\beta_{\dot\alpha}\over x^2} \,\zeta_\beta^A(x)\,.
\end{align}
Notice that the inversions change the chirality of Lorentz indices.

We can use explicit expressions for various fields (gauge field, scalar and gaugino) to verify that, up to
compensating gauge transformations, they transform under inversions as conformal primary fields. For instance,
the gauge field \re{A-exp} transforms as
\begin{align}\label{A-U}
I[A^{\dot\alpha\beta}] = x_{\dot\gamma}^\alpha x_\gamma^{\dot\beta} \lr{ U^\dagger  A^{\dot\gamma\gamma} U + U^\dagger  \partial^{\dot\gamma\gamma} U }\,,
\end{align}
where the $SU(2)$ matrices are multiplied according to \re{AB} and the compensating gauge transformation is
\begin{align}\label{U}
U_i{}^{i'} =\delta_i^\alpha\delta^{i'}_{\dot\alpha} {x_{\alpha}^{\dot\alpha}\over (x^2)^{1/2}}\,,\qqqquad
(U^\dagger)_{j'}{}^j  = \delta^j_\alpha\delta_{j'}^{\dot\alpha} {x_{\dot\alpha}^\alpha\over (x^2)^{1/2}}\,.
\end{align}
Here the product of Kronecker delta-functions identifies the $SU(2)$ indices of $U$ with the  Lorentz indices of $x$.
Replacing the gauge field in \re{A-U} with its general expression \re{A-exp}, we find that the lowest component
$A^{(0)}$ satisfies the same relation \re{A-U} whereas for higher components we have 
$I[A^{(n),\dot\alpha\beta}] = x_{\dot\gamma}^\alpha x_\gamma^{\dot\beta}\, U^\dagger  A^{(n),\dot\gamma\gamma} U$.
Applying \re{A-U} we obtain that the light-like Wilson line transforms under the inversions as
\begin{align}\label{E-inv}
I[ E(x_1,x_2)] = U^\dagger(x_1) E(x_1,x_2) U(x_2)\,.
\end{align}
 An immediate consequence of this relation is that the light-like Wilson loop  \re{W4-prod} is invariant under the
 conformal transformations
\begin{align}\label{inv-W4}
I[W_4]=W_4\,.
\end{align}
Substituting \re{E-dec} into \re{E-inv} we find that all components $E^{(4n)}(x_1,x_2)$ of the Wilson line    \re{E-dec}
have to satisfy \re{E-inv}. For the lowest component \re{E0}, this can be verified by direct calculation. For higher components, 
we can use \re{E-inv} to argue that  they have the general form \re{E8}.

To show this, we introduce a pair of  two-dimensional $SU(2)$ vectors 
\begin{align}\label{vec1}
\ket{\tilde \zeta_1^A} = \lr{{\rho \over x_{10}^2+\rho^2}}^{1/2}  \zeta_i^A(x_1)  \,,\qqqquad
E^{(0)}(x_1,x_2)  \ket{\tilde\zeta_2^A}\,,
\end{align}
where $\ket{\tilde\zeta_2^A}$ is obtained from $\ket{\tilde\zeta_1^A}$ by replacing $x_1$ with $x_2$. Making use of \re{zeta-inv} and \re{U} we verify that  both vectors 
transform under the inversions in the same way, e.g.
$
I[\ket{\tilde\zeta_1^A}] =  U^\dagger(x_1)\ket{\tilde\zeta_1^A}\,.
$
In the similar manner, we can show that the vectors
\begin{align}\label{vec2}
\bra{\tilde\zeta_2^B} =  \lr{{\rho \over x_{20}^2+\rho^2}}^{1/2} \zeta^{j B}(x_2)  \,,\qqqquad
\bra{\tilde\zeta_1^B} E^{(0)}(x_1,x_2)  
\end{align}
transform covariantly under inversions with the same weight, e.g. $I[\bra{\tilde\zeta_2^B}]= \bra{\tilde\zeta_2^B}U(x_2)$. 
Then, taking the tensor product of vectors \re{vec1} and \re{vec2}, we can define four $2\times 2$
matrices each satisfying \re{E-inv}. These matrices form the basis over which 
$E^{(4n)}(x_1,x_2)$ can be expanded. The corresponding expansion coefficients depend on the
conformal invariants. The latter are given by the scalar product of vectors of the form \re{zeta-zeta}.
Notice that expressions \re{vec1} and \re{vec2} simplify in the conformal gauge \re{c-gauge}, e.g.
$\ket{\tilde \zeta_1^A} = \rho^{-1/2} \ket{\zeta_1^A} $ and $\ket{\tilde \zeta_2^A} = \rho^{-1/2} \ket{\zeta_2^A}$.
For the component $E^{(8)}(x_1,x_2)$, this leads to \re{E8}.

Let us show that  the two top components of the Wilson line vanish, 
$E^{(12)}(x_1,x_2)=E^{(16)}(x_1,x_2)=0$ (see Eq.~\re{E12}). The top component 
is proportional to the product of all fermion modes $E^{(16)}(x_1,x_2)\sim\prod_{\alpha, A}\zeta_{1,\alpha}^A\zeta_{2,\alpha}^A$. Since $\vev{p_1\zeta_1} = \vev{p_1\zeta_2}$ for $x_{12} = |p_1]\bra{p_1}$ (see \re{zeta12}), we have $E^{(16)}(x_1,x_2) \sim \prod_A\vev{p_1 \zeta_1^A}^2 =0$. The component $E^{(12)}(x_1,x_2)$ has the form similar to \re{E8} with $P_{CD},\dots,S_{CD}$ being homogenous polynomials of degree $5$ in variables \re{zeta-zeta}.
As we will see in a moment, these variables are proportional to $\vev{p_1\zeta_1^A}$. 
Then, $P_{CD},\dots,S_{CD}$, being homogenous polynomials in $\vev{p_1\zeta_1^A}$
of degree $5$, have to vanish since they are necessarily proportional to the square of a fermion mode leading to 
$E^{(12)}(x_1,x_2) \sim \vev{p_1 \zeta_1^A}^2 =0$. Indeed,
let us choose $\ket{\zeta_1^A}\sim \ket{p_1}$. It follows from $\zeta_1-\zeta_2=x_{12}\bar\eta$ that $\ket{\zeta_2^A}\sim \ket{p_1}$.
The first  two expressions in \re{zeta-zeta} obviously vanish in this case whereas the last one reduces to
\begin{align}
 \bra{p_1} E^{(0)}(x_1,x_2)\ket{p_1} = {1\over\rho^2}\bra{p_1}{x_{10}x_{20}}\ket{p_1}= - {1\over\rho^2}\bra{p_1}{x_{10}x_{12}}\ket{p_1} =0\,.
\end{align}
Here in the first relation we used \re{E0-gauge} and in the second one replaced $x_{20}=x_{10}-x_{12}$.
Thus, all expressions in \re{zeta-zeta} vanish for $\ket{\zeta_1^A}\sim \ket{p_1}$ and, therefore, they are proportional 
to $\vev{p_1\zeta_1^A}$.

\section{$D-$functions in the light-cone limit}\label{app:D-fun} 

The integral over collective coordinates of instantons can be expressed in terms of $D-$functions.  For $x_{ij}^2\neq 0$ they are defined as \cite{DHoker:1999kzh}
\begin{align}\label{D-4} 
D_{\Delta_1\Delta_2\Delta_3\Delta_4} {}&= \int d^4 x_0 \int {d\rho\over\rho^5} \prod_i \lr{\rho\over x_{i0}^2+\rho^2}^{\Delta_i}\,.
\end{align}
For our purposes it is sufficient to consider the special case $\Delta_1=\Delta_3$ and $\Delta_2=\Delta_4$.
Defining 
\begin{align}\label{barD-4}
\bar D_{\Delta_1\Delta_2} = (x_{13}^2)^{\Delta_1}
(x_{24}^2)^{\Delta_2}D_{\Delta_1\Delta_2\Delta_1\Delta_2}\,,
\end{align}
we find that, in virtue of conformal symmetry, $\bar D_{\Delta_1\Delta_2}$ only depends on the cross-ratios $u$ and $v$.
It admits the Mellin integral representation \cite{Arutyunov:2000ku}
\begin{align}\label{bar-D}
\bar D_{\Delta_1\Delta_2} = K \int_{-\delta-i\infty}^{-\delta+i\infty}{ dj_1 dj_2\over (2\pi i)^2} u^{j_1} v^{j_2}
\Gamma(
j_1+j_2+\Delta_1)
\Gamma( j_1+j_2 +\Delta_2) \left[\Gamma(-j_1)\Gamma(-j_2)\right]^2 \,,
\end{align}
where $K = \pi^2 {\Gamma\lr{ \Delta_1+\Delta_2-2} /[2\Gamma^2(\Delta_1)\Gamma^2(\Delta_2)]}$ and
integration goes along imaginary axis slightly to the left from the origin, $0< \delta <1$.

Let us examine \re{bar-D} in the light-cone limit $x_{i,i+1}^2\to 0$, or equivalently $u,v\to 0$. In this limit, the $\bar D-$function develops
logarithmic singularities in $u$ and $v$. Indeed, closing the integration contour in \re{bar-D} to
the right half-plane and picking up the residue at $j_1=j_2=0$ we find
\begin{align}\label{D-bare}
 \bar D_{\Delta_1\Delta_2}={\pi^2\over 2}  {\Gamma\lr{ \Delta_1+\Delta_2-2} \over \Gamma(\Delta_1)\Gamma(\Delta_2)}\Big[\ln u \ln v
 + (\ln u + \ln v)(C_1+C_2-2\psi(1))\Big] + \dots\,,
\end{align}
where $C_i = \psi(\Delta_i)$ is expressed in terms of Euler $\psi-$function. Here dots denote terms
suppressed by powers of $u$ and $v$.
 
Computing instanton corrections to light-like Wilson loop we encounter the same integral \re{barD-4}
but evaluated for $x_{i,i+1}^2=0$. To regularize its divergences we modify the integration measure
in \re{D-4} following \re{dx0} and arrive at $\bar D^{(\epsilon)}_{\Delta_1\Delta_2}$ defined in \re{d-eps}. We expect that $\ln u\ln v$ singularity of \re{D-bare} should 
translate into a double pole, $\bar D^{(\epsilon)}_{\Delta_1\Delta_2}\sim 1/\epsilon^2$. Performing integration in \re{d-eps} we obtain
\begin{align} 
\bar D^{(\epsilon)}_{\Delta_1\Delta_2} {}&= 2K  (x_{13}^2)^{\Delta_1}
(x_{24}^2)^{\Delta_2} \mu^{-2\epsilon} \int_0^\infty  \, {\prod_{i=1}^4 dt_i \, t_i^{\Delta_i-1}  }\big(\sum_i t_i\big)^{2\epsilon}
\e^{-x_{13}^2 t_1 t_3 -x_{24}^2 t_2 t_4}\,,
\end{align}
with $\Delta_3=\Delta_1$ and $\Delta_4=\Delta_2$.
For small $\epsilon$ the calculation of this integral yields
\begin{align} \notag\label{D-eps-res}
 \bar D^{(\epsilon)}_{\Delta_1\Delta_2} {}& = \pi^2  {\Gamma\lr{ \Delta_1+\Delta_2-2} \over \Gamma(\Delta_1)\Gamma(\Delta_2)}\bigg[{1\over \epsilon^2}  (\mu^2 x_{13}^2)^{-\epsilon} +{1\over \epsilon^2}  (\mu^2 x_{24}^2)^{-\epsilon} 
 \\
 {}&
 +{C_2\over \epsilon}  (\mu^2 x_{13}^2)^{-\epsilon} +{C_1\over \epsilon}  (\mu^2 x_{24}^2)^{-\epsilon} 
-\frac12\ln^2(x_{13}^2/x_{24}^2) +C_1C_2-{2\over 3}\pi^2+ O(\epsilon) \bigg]\,,
\end{align} 
with $C_i$ the same as in \re{D-bare}.

 We verify that $\bar D^{(\epsilon)}_{\Delta_1\Delta_2}$ satisfies the evolution equation
\begin{align}
\lr{\mu^2 {\partial\over \partial\mu^2}}^2 \bar D^{(\epsilon)}_{\Delta_1\Delta_2} = 2\pi^2  {\Gamma\lr{ \Delta_1+\Delta_2-2} \over \Gamma(\Delta_1)\Gamma(\Delta_2)}\,,
\end{align}
which should be compared with \re{Z}. 
In distinction from \re{D-bare}, the conformal invariance of $\bar D^{(\epsilon)}_{\Delta_1\Delta_2}$ is
broken by light-cone singularities that  appear as poles in $\epsilon$. We notice that the poles
depend on one of the kinematical variables,  $x_{13}^2$ or $x_{24}^2$, and, therefore, they do not
contribute to the mixed derivative of $\bar D^{(\epsilon)}_{\Delta_1\Delta_2}$ with respect to both variables.
Applying this derivative to \re{D-bare} and  \re{D-eps-res} we arrive at the same result
\begin{align}
{\partial \over \partial \ln x_{13}^2} {\partial \over \partial \ln x_{24}^2} \bar D_{\Delta_1\Delta_2} ={\partial \over \partial \ln x_{13}^2} {\partial \over \partial \ln x_{24}^2} \bar D^{(\epsilon)}_{\Delta_1\Delta_2} =\pi^2  {\Gamma\lr{ \Delta_1+\Delta_2-2} \over \Gamma(\Delta_1)\Gamma(\Delta_2)}\,.
\end{align}
This relation implies that conformal anomaly of $\bar D^{(\epsilon)}_{\Delta_1\Delta_2}$ is annihilated by
the mixed derivative.
 
\section{Gauge field in the instanton background}\label{app:gauge}
       
In this appendix, we present explicit expressions for different components of the gauge field \re{A-exp}.
The lowest component $A^{(0)}$ is given by the BPST instanton \re{BPST}. The remaining
components $A^{(4)}$  and $A^{(8)}$ were derived in \cite{Alday:2016jeo}
\begin{align} \label{A48}
\notag
{}&  A^{(4)}_{\alpha\dot\alpha}   
= - \frac1{12}\epsilon_{ABCD} \zeta_\alpha^{A} \zeta^{\beta B} (\zeta^C  D_{\beta\dot\alpha} F\zeta^D) 
- \frac12\epsilon_{ABCD}  \zeta_\alpha^{A} \bar\eta_{\dot\alpha}^B (\zeta^C F\zeta^D) \,,
\\[2mm]
{}& A^{(8)}_{\alpha\dot\alpha} = -\frac32\zeta^8 \big[ D_{\alpha\dot\alpha}   F^{\beta\gamma} ,F_{\beta\gamma} \big]   - \frac32 (\zeta^7)_{\beta A}
\bar\eta^A_{\dot\alpha} \big[F^{\beta\gamma}, F_{\gamma\alpha} \big]\,,
\end{align}       
where $D_{\alpha\dot\alpha} = \partial_{\alpha\dot\alpha} + A^{(0)}_{\alpha\dot\alpha} $ and
$F_{\alpha\beta} =  \epsilon^{\dot\alpha\dot\beta} D_{(\alpha\dot\alpha} D_{\beta)\dot\beta}$ is a self-dual (chiral) part of the gauge strength tensor of the BPST instanton
\begin{align}
F_{\alpha\beta,ij} {}&= -8\rho^2 {\epsilon_{i\alpha}\epsilon_{j\beta} +  \epsilon_{j\alpha}\epsilon_{i\beta} \over [(x-x_0)^2 + \rho^2]^2} \,.
\end{align}
The Grassmann variable $\zeta_\alpha^A(x)$ is given by a linear combination of fermion modes
\begin{align}\label{xi}
\zeta_\alpha^A(x) = \xi_\alpha^A + x_{\alpha\dot\alpha}\bar\eta^{\dot\alpha A}\,,
\end{align}       
and the following notations are used
\begin{align}
(\zeta^C F\zeta^D) = \zeta^{\alpha C} F_{\alpha\beta} \zeta^{\beta D}\,,\qqquad
(\zeta^7)_{\beta A} = { \partial \zeta^8 / \partial \zeta^{\beta A}}\,,\qqquad
\zeta^8=\prod_{\alpha A} \zeta^{\alpha A} \,.
\end{align}
Notice that the second term on the right-hand side of \re{A48} depends on $\bar\eta$.
As explained in \cite{Alday:2016jeo}, its form is uniquely fixed by the conformal symmetry. Namely, it can be 
determined from the requirement for \re{A48} to satisfy \re{A-U}.       

To compute the Wilson line \re{Es}, we need the expressions for the projection of the gauge field on the  edges 
of the light-like rectangle. Let us consider the segment $[x_1,x_2]$ and define the $SU(2)$ matrix
\begin{align}\label{calA}
\mathcal A^{(n)}(t) = i \dot x^\mu (t) A^{(n)}_\mu(x(t)) = -\frac12 x_{12}^{\dot\alpha\alpha} A^{(n)}_{\alpha\dot\alpha}(x(t)) 
= - \frac12 \bra{p_1}A^{(n)}(x(t))  |p_1]  \,,
\end{align}
where $x(t) = (1-t) x_1 + t x_{2}$ and $x_{12}^{\dot\alpha\alpha}=|p_1]\bra{p_1}$ is a light-like vector. 
To simplify the calculation we use the gauge \re{c-gauge}.
Replacing the gauge field in \re{calA} with \re{BPST} and \re{A48} we find after some algebra
\begin{align}\notag\label{As}
 \mathcal A^{(0)}  {}&= {1\over \rho^2}  x_{10} x_{20} \,,
\\[2mm] \notag
\mathcal A^{(4)}   {}&= {4\over \rho^2}     \epsilon_{ABCD} \ket{\zeta^A(x(t))}  \bra{\zeta_1^{B}} E^{(0)}(x_1,x_2)\ket{\zeta_2^{C}} \bra{\zeta^D(x(t))}\,,
\\[2mm] \notag
\mathcal A^{(8)}  {}& = {8\over 15\rho^4}\zeta^6_{AB}(x(t))  \Big[ E^{(0)}(x(t),x_1) \ket{ {\zeta_1^{A}}} \bra{\zeta_2^{B}} E^{(0)}(x_2,x(t))
\\ 
{}& \hspace*{28mm} - E^{(0)}(x(t),x_2)\ket{ {\zeta_2^{A}}} \bra{\zeta_1^{B}}E^{(0)}(x_1,x(t)) \Big]\,,
\end{align}
where 
$ \zeta^6 _{AB} = \epsilon_{ACDE} \epsilon_{BC'D'E'} (\zeta^2)^{CC'} (\zeta^2)^{DD'} (\zeta^2)^{EE'}$ with $(\zeta^2)^{AB}= \zeta^{\alpha A}\zeta_\alpha^B$ and  we used notation for the $SU(2)$ matrices $ \ket{ {\zeta_1^{A}}} \bra{\zeta_2^{B}} \equiv \zeta_{1\,i}^{A}\,\zeta_2^{B\,j}$.

The dependence on fermion modes $\xi_\alpha^A$ and $\bar\eta^{\dot\alpha A}$ enters into \re{As} through linear
combination \re{xi}
\begin{align}\label{zeta}
\zeta_\alpha^A(x(t)) 
=  (1-t)\, \zeta_{1\alpha}^A
+ t \, \zeta_{2\alpha}^A\,.
\end{align}
Its value at the end points is denoted as
$\zeta(x_1)=\zeta_1$ and $\zeta(x_2)=\zeta_2$. Since $(\zeta_1 - \zeta_2)_\alpha^A = (x_{12})_{\alpha\dot\alpha}\bar\eta^{\dot\alpha A}$, these variables 
satisfy the relation $(x_{12})^{\dot\alpha\alpha}(\zeta_1 - \zeta_2)_\alpha^A = 0$ or equivalently
\begin{align}\label{zeta12}
\vev{p_1 \zeta_1^A} -\vev{p_1 \zeta_2^A}=0\,,
\end{align}
where spinor $\ket{p_1}$ defines the light-like vector 
$x_{12}^{\dot\alpha\alpha} = |p_1]\bra{p_1}$. 
  
We would like to stress that the relations \re{As} hold in the conformal gauge \re{c-gauge}.  
The light-like Wilson line \re{E0} is given in this gauge by the following expression 
\begin{align}\label{E0-gauge}
E^{(0)}(x_1,x_2) = 1+ {x_{10}x_{20}\over\rho^2} \,.
\end{align}
The Wilson lines entering the expression for $\mathcal A^{(8)}(t)$ in \re{As} are linear functions of $t$ 
\begin{align}\notag\label{E0-gauge1}
{}& E^{(0)}(x_1,x(t)) = 1+ t {x_{10}x_{20}\over\rho^2}\,,\qqqquad E^{(0)}(x(t),x_2) = 1+(1- t) {x_{10}x_{20}\over\rho^2}\,,
\\
{}& E^{(0)}(x(t),x_1) = 1- t {x_{10}x_{20}\over\rho^2}\,, \qqqquad E^{(0)}(x_2,x(t)) = 1-(1- t) {x_{10}x_{20}\over\rho^2}\,.
\end{align}       
We can use relations \re{As} -- \re{E0-gauge1} to compute all components of the light-like Wilson line  \re{Es} in the conformal 
gauge \re{c-gauge}. The lowest component is given by \re{E0-gauge}. For the remaining components we find after some algebra   
relations \re{E4} and \re{E8}.    

Comparing  \re{E0-gauge} with the general covariant expression \re{E0} we observe that 
$E^{(0)}(x_1,x_2)$ is polynomial in $x_i$ in the conformal gauge. Examining the relations  \re{E4} and \re{E8}
it is easy to see that the same is true for higher components of the light-like Wilson line. As a consequence, all components  
of the Wilson loop \re{Ws} are also polynomial in $x_i$ in the conformal gauge. This leads to important consequences for the 
function $f_R(I_{13},I_{24})$ defined in \re{F_R}. It depends on the conformal invariants \re{Inv} which are given 
in the gauge \re{c-gauge} by
\begin{align} 
I_{13} = {x_{13}^2\over \rho^2}\,,\qqqquad
I_{24} = {x_{24}^2\over \rho^2}\,.
\end{align}
The fact that the expression on the right-hand side of \re{f-gen} is polynomial 
in $x_i$ in the conformal gauge implies that $f_R(I_{13},I_{24})$ is in fact polynomial in $I_{13}$ and $I_{24}$. In other words, the sum over $\ell_1$ and $\ell_2$
in \re{f-gen}, \re{W-sum} and \re{cusp-f} actually contains a finite number of terms. 
       
\bibliographystyle{JHEP} 


\providecommand{\href}[2]{#2}\begingroup\raggedright\endgroup       
       
\end{document}